\documentclass[useAMS,usenatbib,fleqn]{mnras}
\usepackage{amsmath}
\usepackage{multirow}
\usepackage{graphicx}
\usepackage{color,soul}
\usepackage{epsfig}
\usepackage{amssymb,amsmath}
\usepackage{aas_macros}
\usepackage{fixltx2e}
\defcitealias{Ricarte&Natarajan2018}{RN18}

\title[Signatures of Seeding]{
The Observational Signatures of Supermassive Black Hole Seeds
}

\author[Ricarte \& Natarajan]{Angelo Ricarte$^1$,
Priyamvada Natarajan$^1$ \\
$^1$ Department of Astronomy, Yale University, 52 Hillhouse Avenue, New Haven, CT 06511 \\
}

\date{\today}

\begin{document}
\pagerange{\pageref{firstpage}--\pageref{lastpage}} \pubyear{2017}
\maketitle

\begin{abstract}
The origin and properties of the initial black hole seeds that grow to produce the observed population of accreting sources remain to be determined.  It is a challenge to uniquely disentangle signatures of seeding from fueling and dynamics that shapes the assembly history of growing black holes.  To address this, we use a semi-analytic model developed to track the growth of supermassive black holes adopting multiple prescriptions for accretion.  In contrast with earlier treatments, we explore the interplay between seeding models and two accretion modes.  We find that signatures of the initial seeding do survive in the following observational probes:  the black hole occupation fraction; contribution to the unresolved X-ray background; low-luminosity and high-redshift luminosity functions; and in gravitational wave event signatures.  We find that the behaviour of the low-mass end of the $M_\bullet-\sigma$ relation is dominated by uncertainties in the adopted accretion prescriptions and does not offer clear discrimination between seeding models. We make concrete predictions for future surveys, particularly for the Lynx X-ray surveyor and LISA (The Laser Interferometer Space Antenna) mission, which will each provide different and yet strong constraints on the seed population.  Black hole coalescences detected by LISA and high-redshift quasar luminosity functions observed by Lynx will offer the sharpest seeding discriminants.  Although the signatures of the black hole seeding mechanism that persist remain linked to our understanding of black hole accretion and dynamics, we offer new insights on how these upcoming multi-wavelength observations could be leveraged to effectively disentangle them.
\end{abstract}

\begin{keywords}
black hole physics --- galaxies: active --- quasars: general
\end{keywords}

\section{Introduction}
\label{sec:introduction}

From a census of the local universe, it appears that every galaxy hosts a supermassive black hole (SMBH) in its centre, and the mass of the black hole correlates with host galaxy properties such as bulge mass and velocity dispersion \citep{Ferrarese&Merritt2000,Tremaine+2002,Kormendy&Ho2013}.  Some of these SMBHs are also detected at the highest redshifts that our telescopes can currently probe requiring them assemble at the earliest epochs.  These SMBH ``monsters'' have masses inferred by broad-line measurements on the order of $\sim 10^9 \ M_\odot$ by redshifts $z\sim6$ \citep{Fan+2003,Mortlock+2011,Wu+2015,Banados+2018}.  How did these behemoths assemble their masses so early on in the universe? Was the growth process remarkably efficient? And what were the masses of the initial seed black holes that could have grown so rapidly? These are pressing open questions about the origin and properties of the first black hole seeds in cosmology today. The continuing discovery of this early massive accreting black hole population strains the traditional model for seed formation, that of the remnants of the first stars that produce light initial seeds, given the short amount of time available to assemble these objects.  Assembling these large black hole masses, this early in the history of the universe, requires a confluence of several optimal conditions for the environment of the initial black hole seeds.  These include a vast gas reservoir and feedback that does not interrupt the gas supply to the active galactic nucleus (AGN).  Such conditions, while available, are difficult to sustain over the required period of time to grow to large final masses from low-mass initial seeds (seeds with masses ranging from 10 - 100 $M_\odot$) \citep{Park+2016,Pacucci+2017}.  As a result, there has been a rich exploration of seeding models and growth mechanisms that could account for the masses of black holes powering these high redshift quasars \citep[see][for reviews]{Volonteri&Bellovary2012,Haiman2013,Natarajan2014}.  Broadly speaking, modellers have focused on discriminating between two classes of seed models:  light and heavy seeds. 

Light seeds are hypothesised to be common and initialised with low masses, on the order of 100 $M_\odot$.  Such low-mass black holes could be the naturally expected remnants of massive Population III (Pop III) stars.  Early work suggested that the Pop III initial mass function (IMF) would be skewed toward high masses \citep{Bromm+2002}.  Yet in more recent simulations, Pop III birth clouds fragment more easily, moving these masses downward \citep{Clark+2011,Greif+2012,Latif+2013,Hirano+2014,Stacy+2016}.  Recent gravitational wave observations of $\sim 30 \ M_\odot$ black hole mergers have also opened speculation that primordial black holes could provide light black hole seeds \citep{Bird+2016}.  Although it is easy to form light seeds, they may have a problem growing to supermassive scales.  They are expected to accrete with lower duty cycles and with a lower Eddington limit, and may struggle to reproduce the observed population of $10^9 \ M_\odot$ SMBHs at $z \sim 6$ \citep{Inayoshi+2016,Pacucci+2017}.

Heavier black hole seeds were originally proposed to ease the timing problem for assembling high black hole masses in such a short amount of time.  In the direct collapse black hole (DCBH) picture, the gas in a protogalactic disk can go dynamically Q-unstable and collect in the centre of a dark matter halo and immediately form a black hole \citep{Bromm&Loeb2003,Lodato&Natarajan2006,Begelman+2006}.  DCBHs, as we report in a recent paper, can grow and account for the observed bright quasars at high redshift and in fact, can also account for the most massive black holes at all epochs.  The first calculation coupling the cosmological context with details of dynamical stability of individual disks to derive an initial mass function for massive initial seeds showed that the special physical conditions required make heavy seeds necessarily causes them to be rarer than their lighter counterparts \citep{Lodato&Natarajan2006,Lodato&Natarajan2007} for the following reasons. Disk instabilities need to carry gas to the centre of the galaxy without fragmenting and forming stars en-route. This process favours dark matter halos with lower angular momentum.  It is also necessary to prevent all cooling channels aside from molecular hydrogen, thereby requiring both metal-free gas and an external radiation field that inhibits H$_2$ formation \citep{Ferrara+2014,Agarwal+2016}.  In addition, the very source of this radiation field must have a sufficiently small tidal gravitational field, so as to not disrupt the halo that would otherwise host a DCBH \citep{Chon+2016}. 

Direct collapse may not be the only way to make heavy seed.  A light seed whose growth is rapidly amplified via early super-Eddington accretion episodes is, as far as our models are concerned, a black hole born as a heavy seed.  For example, an off-centre remnant from a Pop III star cluster could random walk within the cluster and grow its mass substantially up to about $10^{4-5} \ M_\odot$ starting at 10-50 $M_\odot$, over a very short period of time (a few million years) \citep{Alexander&Natarajan2014}.  If such a super-Eddington phase is brief, the accretion is very radiatively inefficient, and these circumstances are rare, then the subsequent evolution of rapidly amplified seeds is indistinguishable from that of DCBHs.  Finally, seeds resulting from the dynamical core collapse of the first star clusters could also create seeds in the heavy or intermediate mass range \citep{Devecchi&Volonteri2009,Stone+2017}. 

In this paper, we explore observational signatures that might help us discriminate between these two seeding models that produce light (and common) versus heavy (and rarer) seeds.  We employ the semi-analytic model (SAM) \citet[][hereafter RN18]{Ricarte&Natarajan2018} that we have developed to trace the growth and evolution of black holes over cosmic time in the context of the larger hierarchical $\Lambda$CDM structure formation model.  In this model, SMBH assembly is linked to the assembly histories of their host dark matter haloes.  In particular, major mergers are assumed to trigger efficient bursts of SMBH growth.  Following the mass build-up of initial black hole seeds over cosmic time, using this framework, we derive the electromagnetic and gravitational wave signatures that may help discriminate between seeding models. We find the observables that contain seeding information lie at the extremes ends of both mass and redshift.  We make concrete predictions for future missions which could uncover the seed populations, particularly the Lynx X-ray mission and LISA experiment.

In earlier work, using similar demographic models to probe the growth history of black hole seeds over redshift, \citet{Volonteri&Natarajan2009} found that the black hole occupation fraction at the low mass end of the local $M_\bullet-\sigma$ relation as well as the scatter in the relation at low black hole masses embed information about the initial seeding. This time-evolved model assumed that black hole growth via accretion occurred at a rate capped at the Eddington rate in episodes triggered by mergers. Our SAM explores a larger range of accretion scenarios and with our more comprehensive exploration of a range of accretion modes and improved modelling of mergers, we now re-examine the signatures of initial seeding that are encoded in observations. To this end, we calculate potential discriminating observational signatures across wavelengths.

The paper is organised as follows: in \S\ref{sec:modelling} we summarise the accretion models implemented in our SAM and define the two seeding prescriptions; in \S\ref{sec:results} we present the signatures of seeding in the context of a variety of observables across multiple wavelengths; in \S\ref{sec:discussion} we discuss the limitations of our modelling prescription; in \S\ref{sec:conclusion} we conclude with a summary of our findings and a discussion of future prospects with the several upcoming and planned space missions: JWST, Lynx and LISA.  Throughout, when required, we adopt values for relevant cosmological parameters from \citet{Planck2016}.  

\section{Our Semi-analytic Model}
\label{sec:modelling}

The semi-analytic model used here to track black hole growth is the one introduced in \citetalias{Ricarte&Natarajan2018}, except here we explore the two seeding scenarios---light and massive seeds.  First, we summarise the key modelling elements presented in \citetalias{Ricarte&Natarajan2018}. Then, we provide the details for the inclusion of two seeding models. We also discuss the key ways in which our models present improvements compared to previous work, principally by making a stronger connection to observations via the adoption of empirical scaling relations, moving away relying solely on dark matter halo properties to trace black hole growth as done earlier. 

\subsection{Model Summary}

In the context of the accepted $\Lambda$CDM paradigm for structure formation, we populate halos with black hole seeds and trace their merger history as well as growth history via accretion.  Monte-Carlo merger trees are generated for $z=0$ halos of 23 different masses ranging from $10^{10.6}$ to $10^{15} \ M_\odot$.  20 trees are calculated for each halo mass to probe cosmic variance.  The resolution evolves with redshift and reaches a resolution of $5 \times 10^6 \ M_\odot$ at $z=20$ \citep{Parkinson+2008}.  

Black holes are seeded in host haloes in the redshift range $15 < z < 20$, when the universe is sufficiently metal-free.  Stepping forward through the merger tree, we assume that black holes merge instantly, {\it but with a 10\% probability,} when their host halos merge, which occurs on the dynamical friction\footnote{ Traditionally, models have immediately removed halos experiencing minor mergers from the bookkeeping.  We now maintain them as potential SMBH merger sites, instead waiting the appropriate dynamical friction time, but shut off steady-mode accretion to mimic ram pressure stripping.} timescale \citep{Boylan-Kolchin+2008}.  It is assumed that the final parsec problem is solved on a shorter timescale than the dynamical friction timescale, due either to gas dynamics or the triaxiality that results from a major merger \citep[see][for a review]{Colpi2014}.  The remaining 90\% of black holes which do not merge are assumed to forever wander the halo and are not further tracked. When a black hole merger occurs, we estimate the gravitational wave recoil \citep{Lousto+2012}, and remove the black hole from the inventory if the recoil exceeds the escape velocity from the centre of the halo.  Rather than simply estimating the escape velocity based on the circular velocity of the dark matter halo, as has been traditionally done, we now employ the model of \citet{Choksi+2017}, which includes a new comprehensive set of physics, including the effect of halo accretion modifying the potential well.  

Within this dynamical model of halo mergers we adopt an accretion model to track the growth of the central black holes. In this work, we implement two independent accretion modes---an episodic burst mode as well as two distinct steady accretion modes, in addition to the seeding and examine the resulting observational signatures.  Whenever a major halo merger occurs, defined as a halo mass ratio of at least 1:10, the burst mode of accretion is triggered.  In this mode, a black hole grows at the Eddington limit until it reaches a cap whose functional form is $M_\mathrm{cap} \propto \sigma^5$, where $\sigma$ is the stellar velocity dispersion of the host.  Such behaviour is expected if feedback occurs in the form of energy-driven winds \citep{King&Pounds2015}.  As discussed in depth in \citetalias{Ricarte&Natarajan2018}, we find it important to accurately estimate $\sigma$ based on galaxy properties instead of dark matter halo properties, as has been traditionally done.  This is necessary to match luminosity functions out to $z=6$ and capture the SMBH ``downsizing'' phenomenon.  We estimate $\sigma$ as a function of redshift and halo mass using two empirically motivated scaling relations: employing a stellar mass-halo mass relation from abundance matching \citep{Moster+2013}, combined with the mass-size relation \citep{Huertas-Company+2013,Mosleh+2013}, which we assume evolves with redshift at high-masses.  In this picture, the black hole swallows gas until its own feedback blows away its fuel supply, depleting the reservoir \citep{Natarajan&Treister2009}.  

It is thought that not all AGN may be triggered by mergers \citep{Mechtley+2016,Villforth+2017}. Hence, if a black hole is {\it not} in the burst mode, it accretes via one of two steady modes, the choice of which is determined at the beginning of the run.  In one set of models, the AGN Main Sequence (MS) models, the black hole accretion rate is set to a thousandth of the star formation rate, as observed in stacked observations of star forming galaxies \citep{Mullaney+2012}.  In the second set of models, the Power Law (PL) models, black holes instead accrete using a universal power-law Eddington ratio distribution that is tuned to roughly reproduce the local luminosity function.  The MS models effectively place a floor on $z=0$ black hole mass of $M_\bullet > 10^{-3} \; M_*$, ensuring the growth of all black holes in low-mass hosts.  The PL models do not consistently grow black holes in low-mass galaxies, but they are tuned to match local luminosity functions.  It is important to note that the MS and PL models are mostly identical except for redshifts $z<2$.  At higher redshifts, black hole accretion is dominated by the merger-triggered burst mode instead of this steady mode.

Finally, when computing mass and luminosity functions, it is important to take into account the accumulated scatter in the scaling relations we have adopted \citep{Somerville2009}.  We inject an additional 0.3 dex scatter in our black hole masses when computing these quantities to reach the level of intrinsic scatter observed in the $M_\bullet-\sigma$ relation \citep{Kormendy&Ho2013}.  In practice, this means that we convolve our mass and luminosity functions with a log-normal distribution of this width.

\subsection{Black Hole Seeds}

Black holes are seeded in the model at redshifts $15<z<20$.  Qualitatively, there are two critical features of  the seeding models:  the initial seed masses, and their abundance.  Increasing the initial mass of a seed makes it easier to assemble $10^9 \ M_\odot$ quasars by $z=6$, and decreases the amount of growth that occurs in the form of luminous accretion. Increasing the abundance of seeds raises the fraction of galaxies which host a SMBH, boosting the contribution of BHs in low-mass galaxies to the total luminous accretion as well as the number of BH-BH mergers.  The heavy seed model was employed in \citetalias{Ricarte&Natarajan2018}.

\subsubsection{Heavy Seeds}

The heavy seeds in our model have a mass function that peaks $10^5 \ M_\odot$.  These can be thought of as the products of direct collapse of gas clouds, or perhaps Pop III remnants which experienced the conditions necessary for rapid growth as outlined in \citet{Volonteri&Rees2005} or \citet{Alexander&Natarajan2014}.  Here, we employ the DCBH model of \citet{Lodato&Natarajan2006,Lodato&Natarajan2007} to determine which halos host seeds, and how to initialise their masses.  In this model, gas in a proto-galactic disc accumulates at the centre of the halo due to non-axisymmetric structures that transport angular momentum out and mass inward.  Assuming that all of this mass accretes onto a central object, the resultant black hole seed mass is given by:

\begin{equation}
M_\bullet = m_dM_h \left[1 - \sqrt{\frac{8 \lambda}{m_d Q_c} \left(\frac{j_d}{m_d}\right) \left(\frac{T_\mathrm{gas}}{T_\mathrm{vir}}\right)^{1/2}} \right],
\end{equation}
\noindent where $m_d$ is the fraction of the halo mass that is in the disc, $j_d$ is the fraction of the halo's angular momentum in the disc, $\lambda$ is the halo spin parameter, $T_\mathrm{gas} = 5000 \ K$ is the gas temperature, $T_\mathrm{vir}$ is the halo's virial temperature, and $Q_c$, the critical Toomre parameter, is a free parameter that we set to 3 to allow for agreement with the local black hole occupation fraction.  $\lambda$ is drawn from a log-normal distribution \citep{Warren+1992}, and a black hole is only seeded in a halo if
\begin{equation}
\lambda < \lambda_\mathrm{max} \equiv m_d Q_c / 8 (m_d/j_d)(T_\mathrm{vir}/T_\mathrm{gas})^{1/2},
\end{equation}
\noindent which ensures that the disk is gravitationally unstable, and
\begin{equation}
T_\mathrm{gas} < T_\mathrm{vir} < T_\mathrm{max} \equiv T_\mathrm{gas} \left(\frac{4\alpha_c}{m_d}\frac{1}{1+M_\bullet/m_dM_h}\right)^{2/3},
\end{equation}
\noindent with $\alpha_c = 0.06$ \citep{Rice+2005}, which ensures that the disk does not fragment into stars.   Compared to \citetalias{Ricarte&Natarajan2018}, we now allow seeds to form in halos which have progenitors in the merger tree.

\subsubsection{Light Seeds}

In the light seeding models, we seed every dark matter halo which is at least a $3.5\sigma$ peak.  Seed masses are drawn from a power law with $dn/d\log M_\bullet \propto M_\bullet ^{-0.3}$, where this shallow slope is derived from simulations of Pop III star formation \citep{Stacy+2016,Hirano+2014}.  This slope is poorly constrained at the present time. This distribution is given a lower limit of 30 $M_\odot$, and an upper limit of 100 $M_\odot$.  The exact details of the Pop III initial mass function such as its cutoffs and slope are rapidly erased by early accretion events.  What matters qualitatively is that these light seeds (i) populate halos more ubiquitously, and (ii) start with much lower masses.

\subsection{Four Model Variants}

To re-iterate, here, we explore two different seeding models and two different steady accretion models which both include the burst mode, for a total of four model variants.  They are named as follows:

\begin{itemize}
\item Light-PL:  Light seeds, power law steady mode.
\item Light-MS:  Light seeds, AGN Main Sequence steady mode.
\item Heavy-PL:  Heavy seeds, power law steady mode.
\item Heavy-MS:  Heavy seeds, AGN Main Sequence steady mode.
\end{itemize}

\section{Results}
\label{sec:results}

Growth via accretion and mergers works to erase the initial seeding conditions, and we seek any distinct signatures that survive, both at high- and low-redshift.  In this section, we systematically explore the observables which may reveal these signatures, and determine whether or not signatures of seeding are separable from signatures of accretion. 

\begin{figure*}
   \centering
   \includegraphics[width=0.8\textwidth]{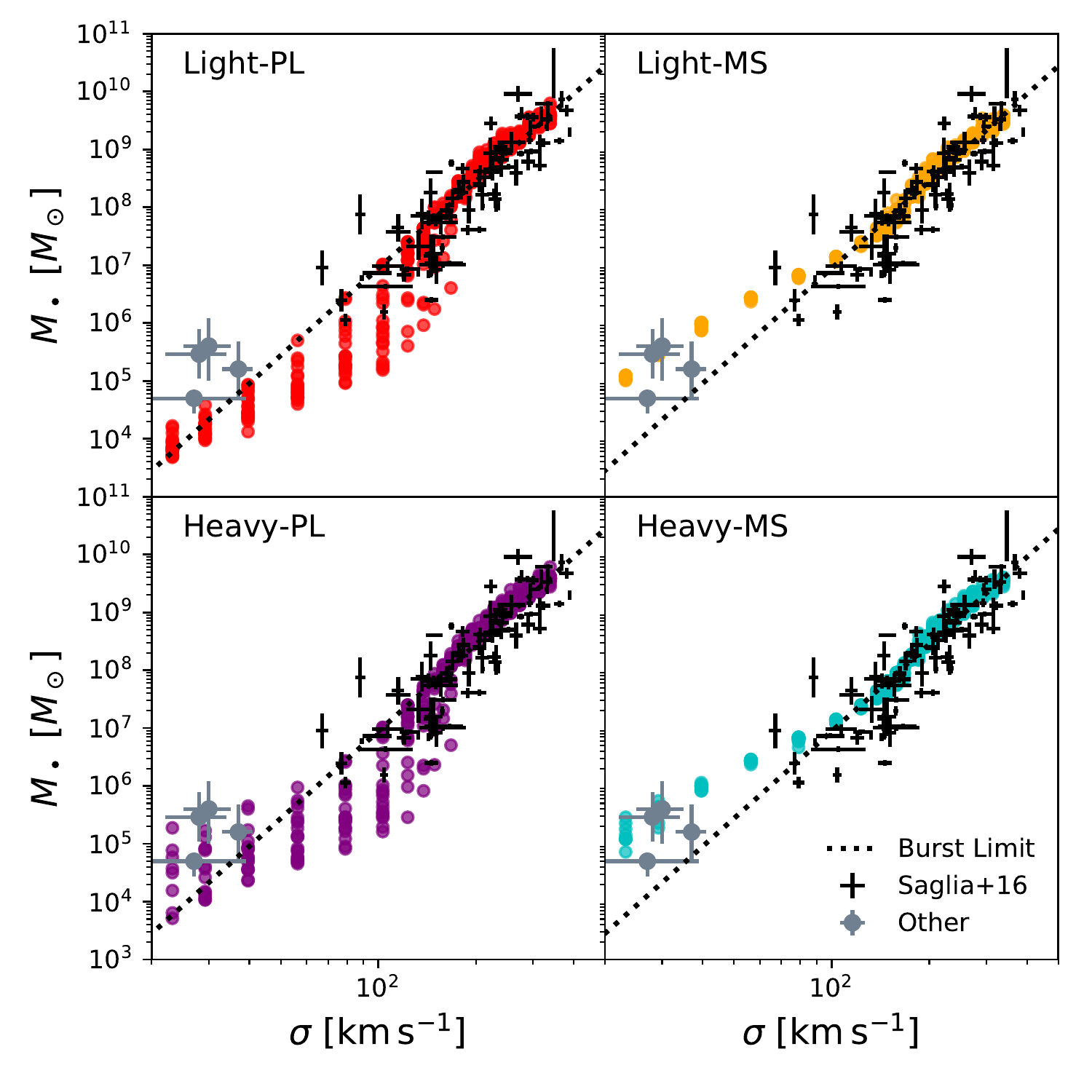}
   \caption{The $M_\bullet-\sigma$ relation at $z=0$ produced by the 4 different SAM models outlined above.  In black, we over-plot a recent compilation of data by \citet{Saglia+2016}, as well as a dotted line which represents the feedback limit during the burst mode of accretion.  In grey, we add both direct and indirect black hole mass measurements in dwarf galaxies RGG 118 \citep{Baldassare+2015}, RGG 119 \citep{Baldassare+2016}, NGC 4395 \citep{DenBrok+2015}, and Pox 52 \citep{Barth+2004}.  Although previous works have argued that the scatter in the low-mass end of this relation is determined and driven by the seeding mechanism, we see no clear difference between seeding models.  Instead, we find that the implementation of the steady mode of black hole accretion determines the level of scatter in the low-mass end of this relation.}
   \label{fig:msigma}
\end{figure*}

\subsection{The $M_\bullet-\sigma$ Relation}

The $M_\bullet-\sigma$ relation is a power-law relationship observed between SMBH mass, $M_\bullet$, and the velocity dispersion of its host's bulge, $\sigma$ \citep[e.g.,][]{Ferrarese&Merritt2000,Tremaine+2002,Kormendy&Ho2013,vandenBosch2016}.  This relationship could be a reflection of the feedback process by which SMBHs shut off their own growth.  For example, it has been shown analytically that momentum-driven wind feedback can naturally produce a $M_\bullet \propto \sigma^4$ relationship, while energy-driven wind feedback can naturally produce a $M_\bullet \propto \sigma^5$ relationship \citep{King2003,Natarajan&Treister2009,King&Pounds2015}.  It is also worth pointing out, however, that the central limit theorem may play a role in establishing these relations \citep{Peng2007,Jahnke&Maccio2011}.

Previous studies have claimed that the behaviour of the low-mass end of the $M_\bullet-\sigma$ relation can distinguish between seeding models.  \citet{Volonteri&Natarajan2009} report that heavy seeds produce a larger scatter at low-velocity dispersion, and that the relation may also flatten at low-mass.  We do not draw the same conclusion, and find instead that the behaviour of the low-velocity dispersion, low black hole mass end is mainly determined by our accretion prescriptions.  We plot the $M_\bullet-\sigma$ relation for each of our models at $z=0$ in Figure \ref{fig:msigma}, the coloured points are derived from the SAM, while points with error bars represent a recent compilation of data from \citet{Saglia+2016}.  Only central galaxies in the SAM are included, since unmodelled environmental processes may alter the stellar content of satellite galaxies and their relationships with their dark matter halos.  As a dotted line, we plot the line which represents our feedback limit during the burst mode, which caps SMBH growth following major mergers.  We also include data points from both dynamical and indirect measurements of black hole masses in a few dwarf galaxies in grey:  RGG 118 \citep{Baldassare+2015}, RGG 119 \citep{Baldassare+2016}, NGC 4395 \citep{DenBrok+2015}, and Pox 52 \citep{Barth+2004}.

In our model, the differences in the scatter at the low-mass end of the $M_\bullet-\sigma$ relation arise mainly due to the different accretion modes (column-to-column) rather than from seeding models (row-to-row).  As discussed in \citetalias{Ricarte&Natarajan2018}, the power law models do not consistently grow low-mass SMBHs to the $M_\bullet-\sigma$ relation.  This is because the power law Eddington ratio distribution is tuned to have a low mean, and low-redshift major mergers would be required to trigger significant black hole growth in low-mass halos.  The AGN Main Sequence models grow black holes in less massive halos to higher mass, since the stellar mass sets a floor on black hole mass in these models.  Note that our adopted feedback cap, which works for high masses, appears to undershoot recent dwarf galaxy black hole masses.  However, these data may not be representative of typical galaxies with velocity dispersions due to selection effects.  Therefore, upcoming data that spans the low mass end more comprehensively is awaited.  On the theoretical end, including more sophisticated and empirically motivated accretion models than currently available might also help break this degeneracy.

We argue that the scatter at the low-mass end of the $M_\bullet-\sigma$ relation is driven primarily not by the seeding model, but rather the process by which these low-mass SMBHs are fuelled.  This was not seen in \citet{Volonteri&Natarajan2009} because their cap on merger-driven black hole growth did not evolve as strongly with redshift at the low-mass end as in our models, and they did not explore the comprehensive range of feasible accretion models studied here (See Figure 3 in \citetalias{Ricarte&Natarajan2018}.)  More information may be retained in the SMBH occupation fraction as claimed by \citet{Volonteri&Natarajan2009}, and we explore this in the next section.

\subsection{The SMBH Occupation Fraction}

\begin{figure}
   \centering
   \includegraphics[width=0.45\textwidth]{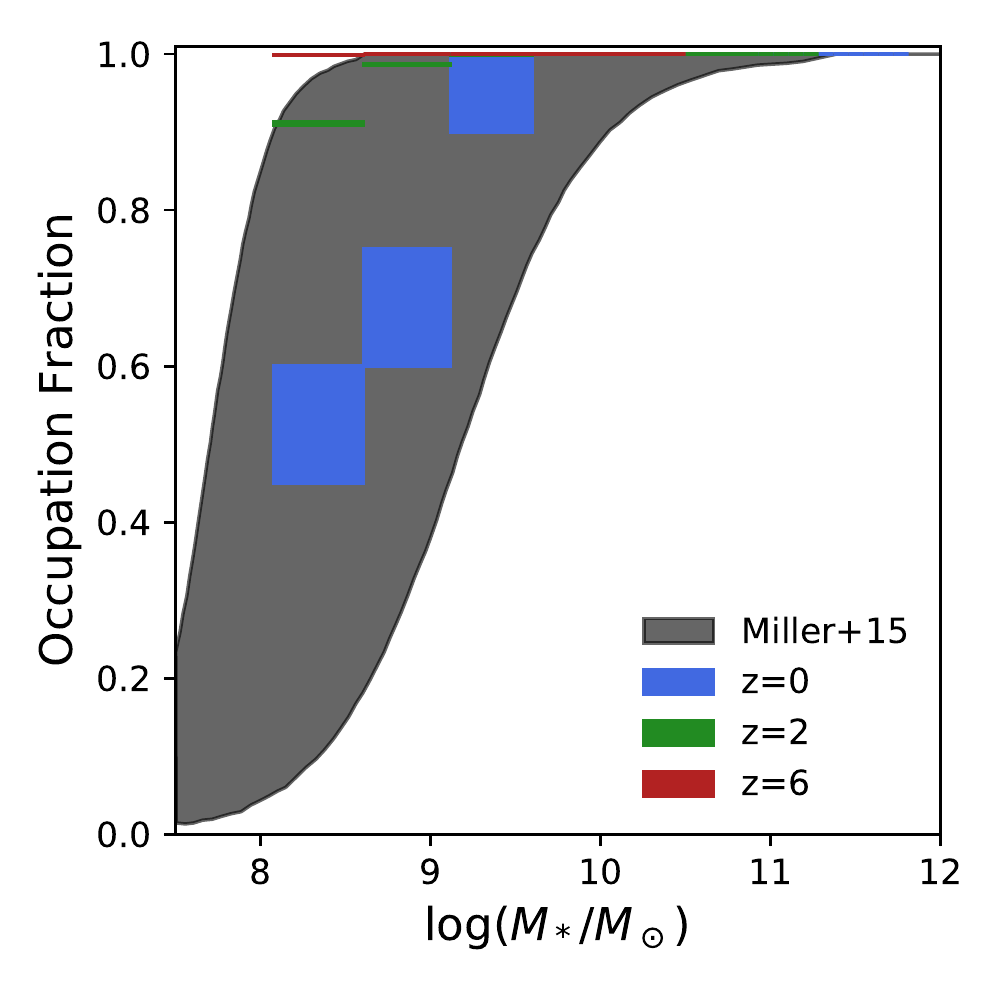}
   \caption{The SMBH occupation fractions derived from our SAM for the heavy seeding model (here, Heavy-MS).  In black, we over-plot estimates based on the X-ray emission from low-mass galaxies \citep{Miller+2015}.  For these stellar masses and redshifts, the occupation fraction of light seeds is unity.}
   \label{fig:occupationFraction}
\end{figure}

Light seeds are assumed to be ubiquitous, while the abundance of heavy seeds is limited by the special conditions required for their formation as noted above. This may leave a detectable signature among galaxies with stellar masses below $\sim 10^{10} \ M_\odot$.  Again, throughout this section, only central galaxies are considered. 

In Figure \ref{fig:occupationFraction}, we plot the SMBH occupation fraction for the halos in one of our heavy seeding models (Heavy-MS) at three different redshifts.  For light seeds, the occupation fraction is unity for all masses and redshifts shown.  We compare our $z=0$ results to the range estimated by \citet{Miller+2015}.  As mentioned in \citetalias{Ricarte&Natarajan2018}, we have set the parameter $Q_c=3$ in order maintain agreement with these observations.  Note that the larger error bars at $z=0$ are the result of having a smaller number of halos  at lower redshift, since our merger trees are built by moving backwards in time.  

For $M_* > 10^{10} \ M_\odot$, the occupation fraction is 1.  Below that mass, the occupation fraction evolves towards higher values with increasing redshift.  This is expected, since any galaxy at $z>0$ represents the progenitor of a galaxy with higher stellar mass at $z=0$.

\begin{figure*}
   \centering
   \includegraphics[width=0.8\textwidth]{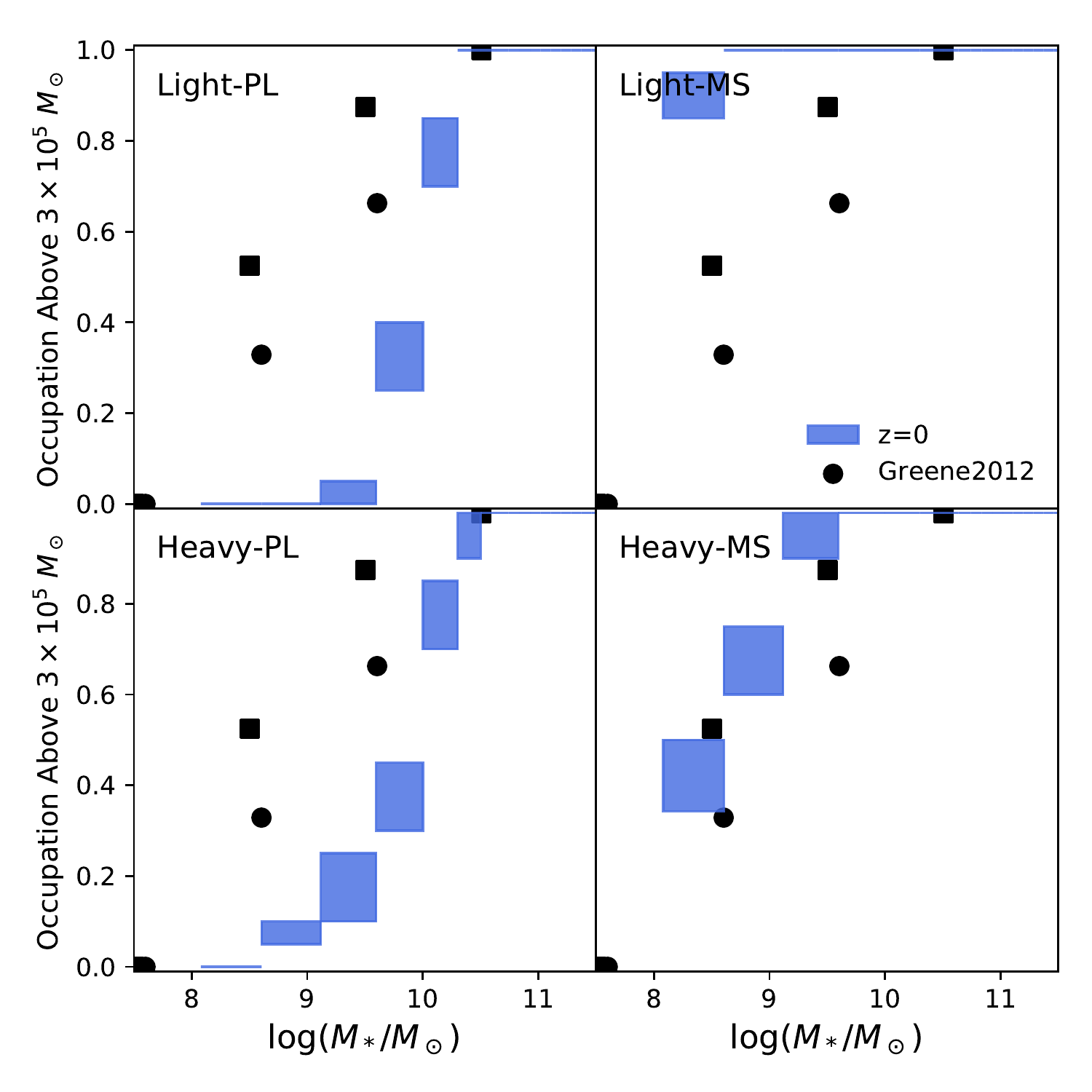}
   \caption{Black hole occupation fractions {\it above $3 \times 10^5 \ M_\odot$}, compared to observational constraints from \citet{Greene2012}.  Our model is shown in blue, while the data from \citet{Greene2012} based on the \citet{Desroches&Ho2009} and \citet{Gallo+2010} samples are displayed as squares and circles respectively.  An occupation fraction defined with a black hole mass threshold depends on how frequently the accretion model pushes existing black holes above this mass.}
   \label{fig:occupationFraction_mt}
\end{figure*}

\subsubsection{Including a Black Hole Mass Threshold} 

\citet{Greene2012} offers a similar observational constraint, the occupation fraction of black holes above a certain mass threshold.  We compare the results from our SAM while making the same mass cut, here $M_\bullet > 3 \times 10^5 \ M_\odot$, in Figure \ref{fig:occupationFraction_mt}.  Occupation fractions estimated from the \citet{Desroches&Ho2009} and \citet{Gallo+2010} are shown as squares and circles respectively, which roughly represents the systematic error of these measurements.  Applying this mass threshold reveals that although the raw occupation fraction may be unity for light seeds, the Light-PL model nevertheless undershoots these observational estimates.  This is due to fact that the PL steady mode does not always grow black holes in low-mass galaxies to their maximum mass.  Although there are significant model uncertainties and systematic uncertainties in the measurements, we comment that Heavy-MS fares the best compared to these constraints.  This model has the low intrinsic occupation fraction of heavy seeds, while also pushing most of them above the mass threshold.

\subsubsection{Weighting by $M_\bullet-\sigma$}

For many of the black holes in our model, we find that their masses are significantly higher or lower than one would expect from the $M_\bullet-\sigma$ relation.  However, there are quantities derived directly from the observational data, by stacking analysis for instance, that may measure a slightly modified version of the statistic of the occupation fraction. From an observational standpoint, a BH under-massive relative to the $M_{\bullet}-\sigma$ relation, is more difficult to detect, and therefore is less likely to contribute to an estimate of the occupation fraction.  To give a concrete example, one might attempt to estimate the occupation fraction by measuring the stacked X-ray luminosity attributed to AGN in galaxies of similar stellar mass, assuming a uniform Eddington ratio distribution along with an $M_\bullet-\sigma$ (or similar) relation.  In this case, it is the mass-weighted occupation fraction of BHs which would be actually measured from the data, since BHs at fixed Eddington ratio would contribute to the stack proportionally to their masses.  We define the mass-weighted occupation of black holes by

\begin{align}
\mathrm{OF}_\mathrm{mass\--weighted}(M_*) = \frac{<M_\bullet>}{M_\sigma},
\end{align}

\noindent where $M_\sigma := M_{\bullet,\mathrm{cap}} = 10^{8.45}  (\sigma/200 \; \mathrm{km}\; \mathrm{s}^{-1})^5 \ M_\odot$.  That is, the mass-weighted occupation fraction is the ratio of a galaxy's black hole mass (possibly zero) to the mass it would have if it lay on the $M_\bullet-\sigma$ relation, for which we use the parameters of our feedback cap.  If every galaxy of a given mass had a SMBH that lay exactly on the $M_\bullet-\sigma$ relation, then the mass-weighted occupation fraction at that mass would be 1.

\begin{figure*}
   \centering
   \includegraphics[width=0.8\textwidth]{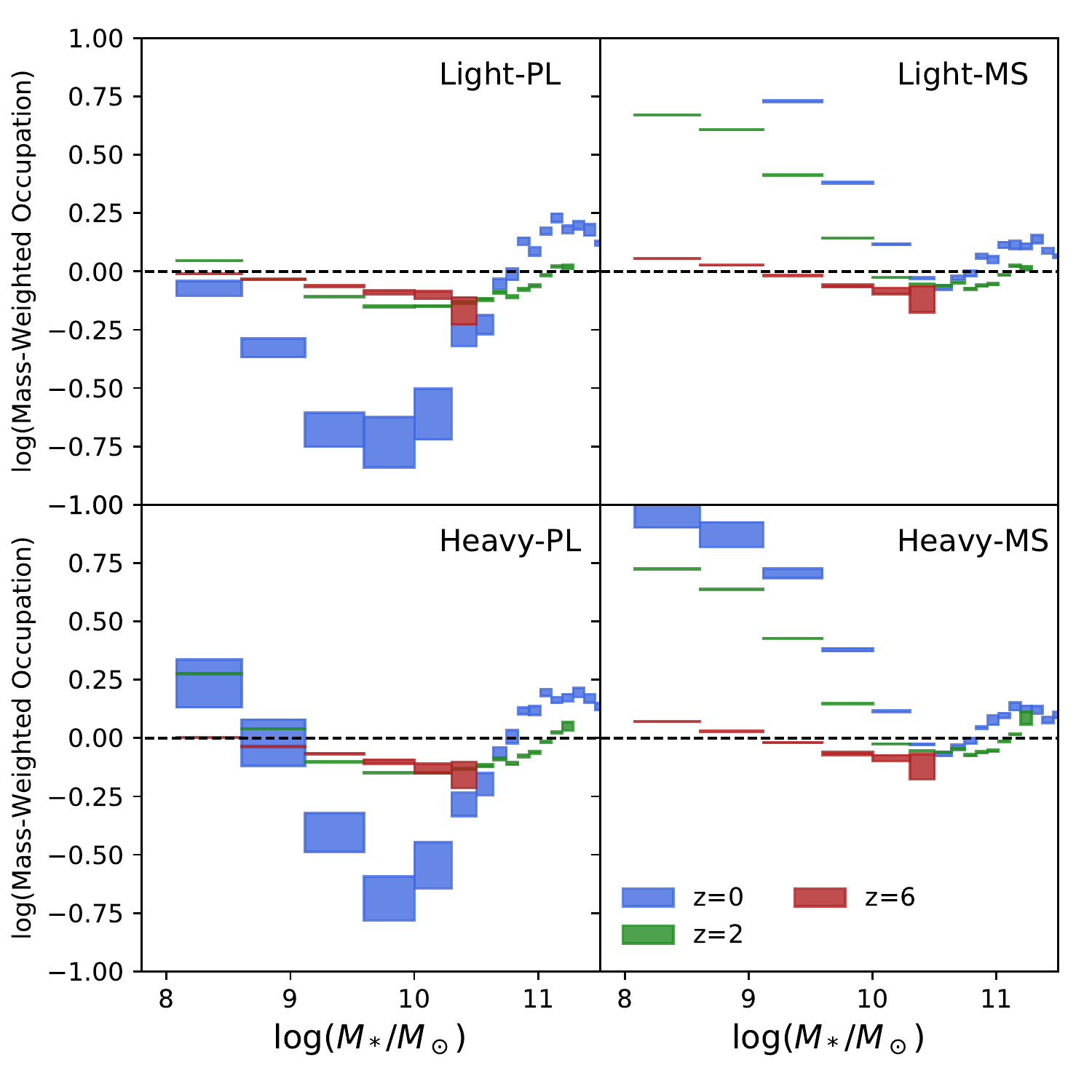}
   \caption{Mass-weighted occupation derived from our SAM, defined as the average black hole mass (possibly zero) divided by the mass expected mass from the $M_\bullet-\sigma$ relation.  At low stellar masses, MS models push black holes above $M_\bullet-\sigma$, while PL models keep black hole masses low.  At high stellar masses, the steady accretion mode pushes black hole masses slightly above our burst mode $M_\bullet-\sigma$ cap.}
   \label{fig:effectiveOccupationFraction}
\end{figure*}

The mass-weighted occupation fraction is plotted in Figure \ref{fig:effectiveOccupationFraction}.  Here, we again see signatures of our accretion models, with the MS models preferentially pushing black holes in low-mass hosts above the $M_\bullet-\sigma$ relation.  At the high-mass end, the steady accretion mode and SMBH mergers push black holes slightly above our burst mode $M_\bullet-\sigma$ cap.  While higher redshifts are shown, we caution that these values are dependent on assumptions about the evolution of the $M_\bullet-\sigma$ relation with redshift, and one's mapping between galaxy properties to $\sigma$.  In our model, galaxy mergers are a more reliable and dominant mode of SMBH growth at higher redshift, our models conform more strongly to the $M_\bullet-\sigma$ relation at these epochs.

\subsection{The High-Redshift Active Population}

For most of cosmic time, redshifts $z < 6$, the properties of the luminous SMBH population are set by host galaxy properties instead of their own initial masses.  In particular, major mergers set the frequency of accretion episodes and the velocity dispersion sets the maximum SMBH mass in a given halo.  Yet closer to the seeding epoch, it may be possible to distinguish models before accretion erases the initial conditions.  For seeding constraints that are independent of accretion models, it is not sufficient to characterise the brightest, most-massive quasars.  The defining measurements will come from low-luminosity, high-redshift sources that are currently just beyond our observational capabilities, but precisely the sources that we expect to detect with upcoming and the next-generation facilities.

\subsubsection{High-Redshift Luminosity Functions}

\begin{figure*}
   \centering
   \includegraphics[width=\textwidth]{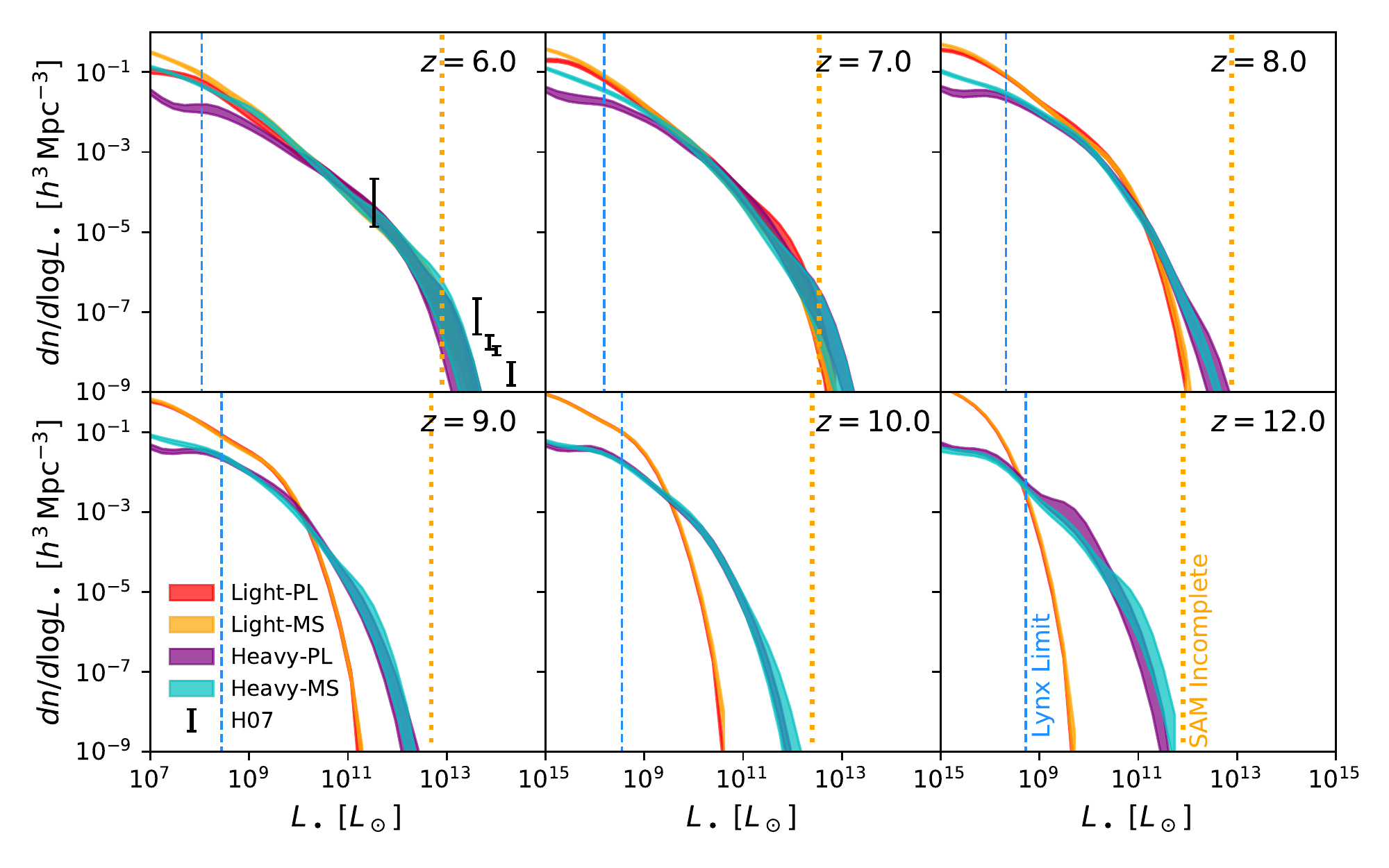}
   \caption{High-redshift luminosity functions predicted by our models.  The luminosity above which we estimate that our SAM is incomplete is marked with the dotted orange line, and the flux limit of a Lynx deep field is shown as a dashed blue line.  At the low-luminosity end, light seeds produce more AGN than the heavy seeds.  The high-luminosity end begins to evolve rapidly for light seeds at $z>9$, since the abundance of high-luminosity AGN is limited by the maximum possible mass that can be attained since the seeding epoch.}
   \label{fig:lum_past6}
\end{figure*}

By $z=6$, although only $\sim 0.1-1\%$ of the total accreted mass density in black holes is accumulated, the signatures of early seeding are already erased from luminosity functions in our model.  This is because by this time, SMBH masses are limited by the properties of their host halos rather than by their own initial masses.  Hence, at all redshifts $z < 6$, the bolometric luminosity functions for light and heavy seeds are identical, and depend primarily on our accretion prescriptions.  In Figure \ref{fig:lum_past6}, we extend luminosity functions beyond our current redshift frontier to $z=12$, using the same procedures as in \citetalias{Ricarte&Natarajan2018}.  The orange dotted line in each panel represents an estimate of the luminosity above which we would be missing at least 50\% of AGN, since our model predicts that they reside in even more biased regions (higher $z=0$ host halo masses) than those probed by our merger trees. This limit therefore, represents the incompleteness limit of our models.  The proposed future X-ray mission concept study Lynx, is set to have 50 times higher sensitivity than the Chandra X-ray telescope and a high angular resolution at 0.5 arcsec or better. The instruments are designed with the goal of bringing into view the faintest accreting sources in the universe from the earliest epochs. A deep field from with the Lynx telescope is planned to reach a flux limit of $10^{-19} \ \mathrm{erg} \; \mathrm{s}^{-1} \; \mathrm{cm}^{-2}$ over an area of 400 arcmin$^{2}$\footnote{https://wwwastro.msfc.nasa.gov/lynx/docs/science/blackholes.html}.  As a dashed blue line, we also plot the sensitivity of a Lynx deep field, converted into a bolometric luminosity using standard cosmology and a 10\% X-ray flux fraction.  In the $z=6$ panel, we include bolometric luminosity function estimates from \citet{Hopkins+2007}.  Here, we search for signatures of seeding at high redshift that therefore do not depend on the steady accretion mode.

To reiterate, the two salient features of the two seed populations are their maximum mass and their abundance. Both of these manifest in these high-redshift luminosity functions.  First, the maximum seed mass determines the evolution of the high-luminosity end of these luminosity functions.  At sufficiently high redshift, the high-luminosity end is limited by the amount of time it takes to grow from the maximum seed mass to the SMBH masses required to produce the observed luminosities at the Eddington rate.  In our model, dictated by the choice of our Pop III IMF, this signature, a rapid evolution in the abundance of high-luminosity AGN, can be seen at $z=9$ and beyond.  Exactly at what redshift the evolution begins depends jointly on the maximum seed mass and the accretion prescription.  Decreasing the seed mass, average Eddington ratio, and duty cycle would all decrease the redshift at which this sudden evolution occurs.  By $z=12$, our two seeding models have very different predictions for the luminosity function.  At these epochs, the detection of even a single source or handful of lensed candidates by JWST might break the impasse in differentiating between seeding models.  

Second, the abundance of seeds manifests in the low-luminosity end of these luminosity functions, at $L_\bullet < 10^{10} \ L_\odot$.  Here, the light seed models predict more low-luminosity AGN than the heavy seed models.  For our model, this is independent of the steady mode for $z>7$, since our merger-triggered burst mode dominates at these redshifts.  However, it is worth noting that recent hydrodynamic simulations suggest that supernova feedback may inhibit accretion in the lowest mass galaxies, which could prevent black holes from accreting in the lowest-mass galaxies \citep{Dubois+2015,McAlpine+2017,Bower+2017}. 

Measuring the luminosity function at these extremes in both luminosity and redshift is currently beyond our observational reach, but the planned Lynx mission would have the specifications and sensitivities to reach these limits and might well uncover the entire population.  We make a quantitative estimates for the Lynx mission by converting our luminosity functions into numbers of discrete sources expected in a Lynx deep field at each epoch.  This conversion applies the following equation,

\begin{align}
\frac{d^2N_\mathrm{obs}}{d \log L_\bullet dz} (z) = \frac{\mathrm{FoV}}{4\pi} \frac{d^2N}{d\log L_\bullet dV} f_\mathrm{obs}(L_\bullet,z) \frac{dV}{dz},
\end{align}

\noindent where $\mathrm{FoV}$ is the survey field of view, $d^2N/d\log L_\bullet dV$ is the luminosity function, and $dV/dz = c d_L^2/(1+z) \cdot dt/dz$ for luminosity distance $d_L(z)$.  The X-ray observable fraction $f_\mathrm{obs}$ is estimated based on $N_H$ distributions measured by \citet{Ueda+2014}, and is on the order of $\sim 0.5$.  We do not account for source confusion.

\begin{figure*}
   \centering
   \includegraphics[width=\textwidth]{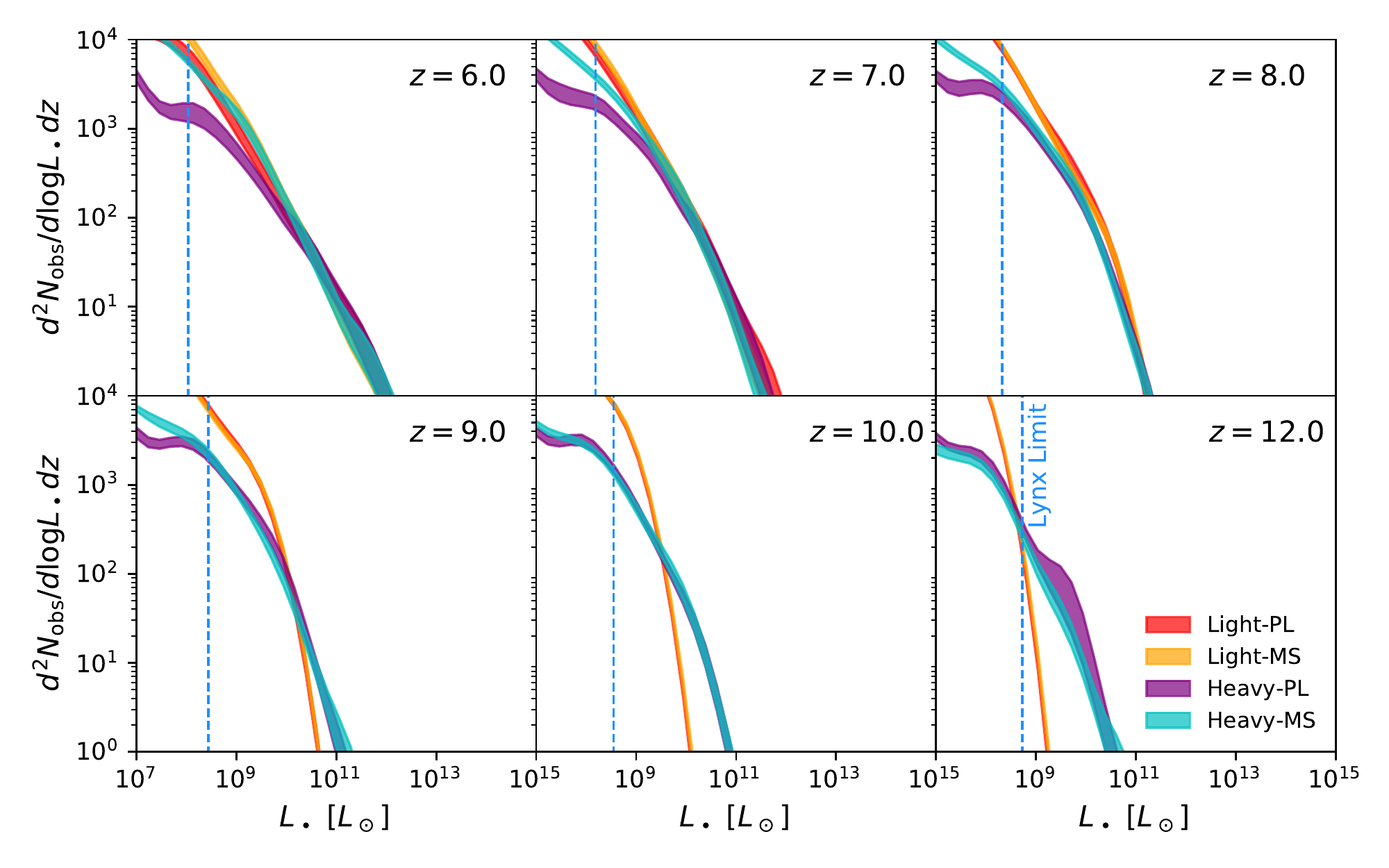}
   \caption{Numbers of objects detectable by Lynx per decade in luminosity per redshift interval.  These are converted from Figure \ref{fig:lum_past6} assuming a 400 arcmin$^2$ field of view.  Remarkably, we predict thousands of high-redshift AGN in a Lynx deep field. Source confusion is not taken into account.}
   \label{fig:lum_past6_lynx}
\end{figure*}

\begin{figure}
   \centering
   \includegraphics[width=0.45\textwidth]{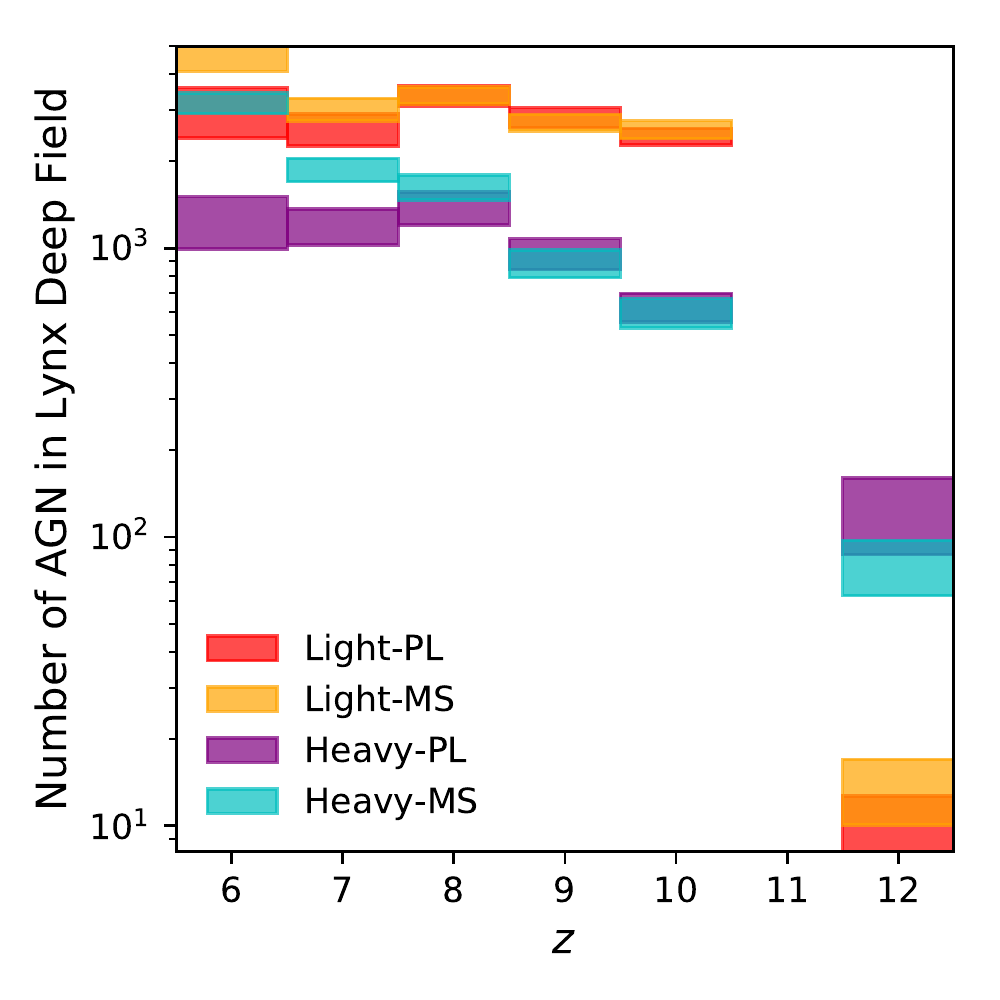}
   \caption{Total numbers of detections in a Lynx deep field for each of our models, also tabulated in Table \ref{tab:lynx}.  Heavy seeds exhibit less redshift evolution than light seeds.  Source confusion is not taken into account.}
   \label{fig:lynx_integrated}
\end{figure}

We provide estimates for $d^2N_\mathrm{obs}/d \log L_\bullet dz$ in Figure \ref{fig:lum_past6_lynx}.  Remarkably, each of our models predicts hundreds to thousands of low-luminosity objects in a Lynx deep field.  Integrating over $d\log L_\bullet$ from the luminosity limit to infinity, we compute the total number of objects that Lynx should detect from each redshift bin.  This is plotted in Figure \ref{fig:lynx_integrated} and listed in Table \ref{tab:lynx}.  These numbers are comparable to the most optimistic scenarios reported in previous work based on evolving scaling relations \citep{Ben-Ami+2018}.  As a sanity check, we perform the same analysis for the Chandra Deep Field-South (CDF-S), assuming a sensitivity of $10^{-17} \ \mathrm{erg} \; \mathrm{s}^{-1} \; \mathrm{cm}^{-2}$ over a $5' \times 5'$ field of view, and obtain on the order of one detection at $5.5<z<6.5$, and zero at subsequent redshifts.  This is consistent with the current estimates for high-redshift X-ray sources in the CDF-S \citep{Cowie+2012}.  Hence, Lynx will be very useful in distinguishing between the two seeding scenarios:  the measured evolution at the low-luminosity end would inform us about the occupation fraction, while the evolution at the high-luminosity end would inform us about the maximum seed black hole mass.

\begin{table*}
\begin{center}
 \begin{tabular}{lcccccc} 
 \hline
Model & $5.5< z < 6.5$ & $6.5<z<7.5$ & $7.5<z<8.5$ & $8.5<z<9.5$ & $9.5<z<10.5$ & $11.5<z<12.5$ \\
\hline
Light-PL & $3000 \pm 600$ & $2600 \pm 300$ & $3400 \pm 300$ & $2800 \pm 200$ & $2400 \pm 160$ & $10 \pm 2$ \\
Light-MS & $4600 \pm 600$ & $3000 \pm 300$ & $3300 \pm 200 $ & $2700 \pm 180$ & $2600 \pm 180$ & $14 \pm 4$ \\
Heavy-PL & $1200 \pm 300$ & $1200 \pm 170$ & $1400 \pm 180$ & $960 \pm 120$ & $630 \pm 70$ & $120 \pm 40$ \\
Heavy-MS & $3200 \pm 300$ & $1900 \pm 170$ & $1600 \pm 160$ & $900 \pm 100$ & $600 \pm 70$ & $80 \pm 17$ \\
 \hline
\end{tabular}
\end{center}
\caption{Expected detections in a Lynx deep field for each of our models.  Thousands of AGN are predicted.  The total number is predicted to evolve more mildly with redshift for heavy seeds than with light seeds.}
\label{tab:lynx}
\end{table*}

\subsubsection{Unresolved Backgrounds}

In addition to X-ray observations of individual sources that can be used to determine the faint-end and bright-end of luminosity functions, these high redshift sources might also been revealed by their integrated emission and therefore contribution to the X-ray background as well as the Infra-red background. The X-ray background contains the cumulative contribution of the population of all accreting black holes. This does not require that we resolve AGN into individual sources.  The flux of the X-ray background can be related to the total amount of matter accreted onto SMBHs at high-redshift using Soltan-like arguments \citep[e.g.,][]{Salvaterra+2012,Cappelluti+2017a}.  This conversion requires assumptions about the radiative efficiencies and computation of the X-ray flux fractions of our AGN sources for which template spectral energy distributions need to adopted.

\begin{figure*}
   \centering
   \includegraphics[width=0.45\textwidth]{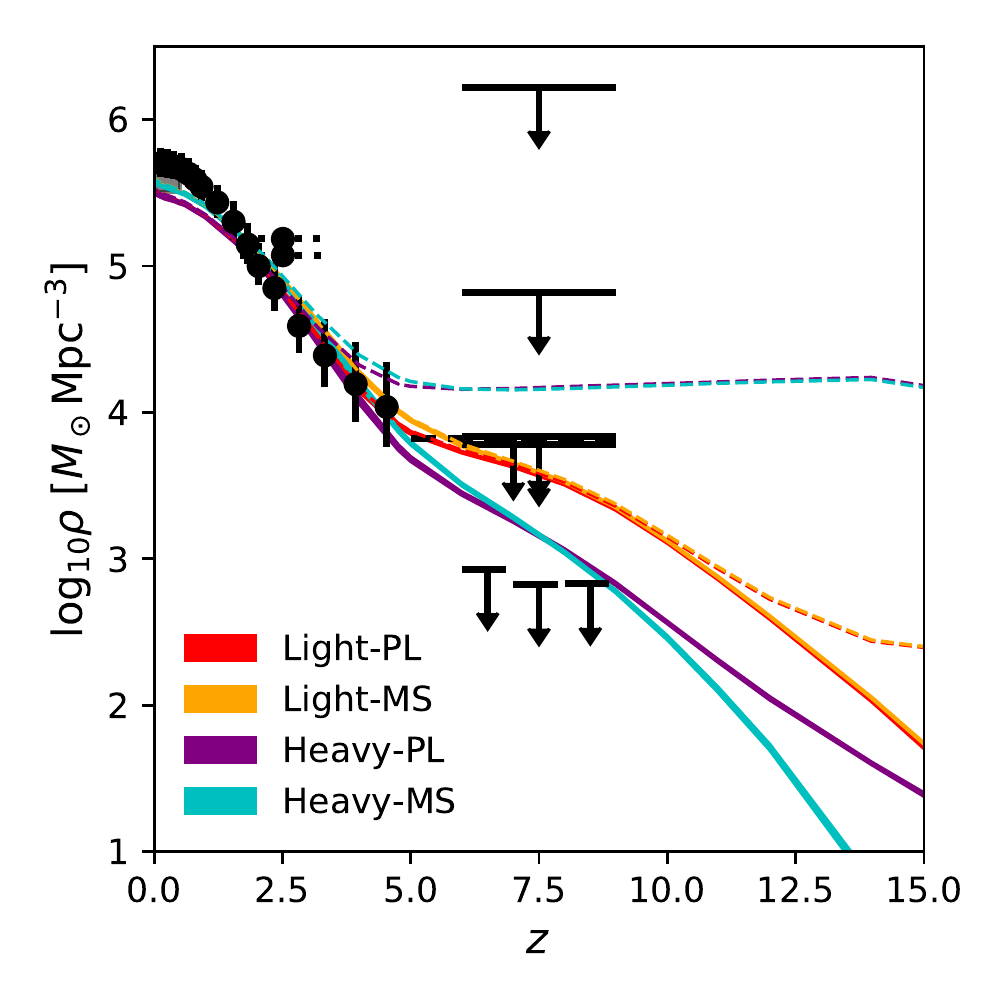}
   \includegraphics[width=0.45\textwidth]{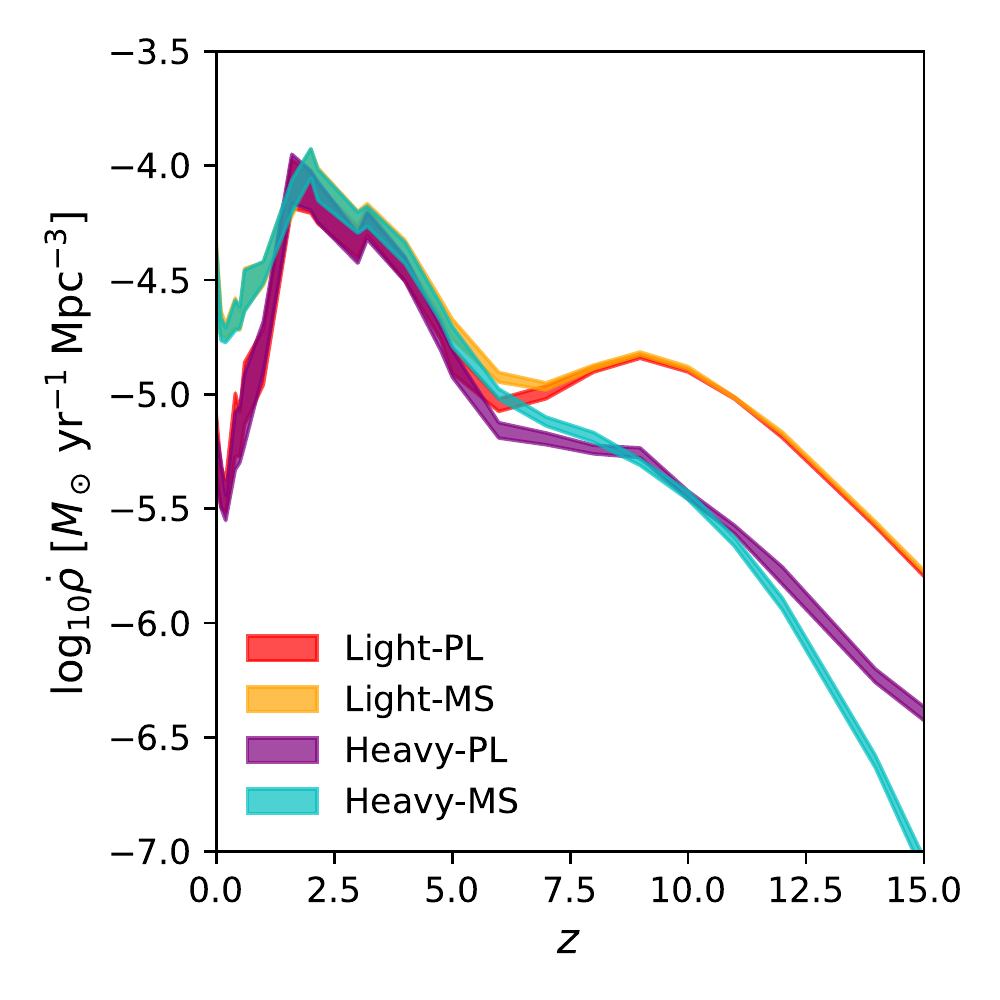}
   \caption{Mass and accretion densities as a function of redshift.  {\it Left}:  The total mass locked up in black holes is shown as dashed lines, while only the contribution due to accretion is shown as solid regions.  Various observational estimates are over-plotted for comparison.  At high-redshift, more mass has been accumulated as the result of accretion in the light seed model.  {\it Right}:  The accretion density onto black holes.  At high-redshift, light seeds must accrete more to catch up to the masses that heavy seeds are initialised with, resulting in a second bump.}
   \label{fig:accretedMassDensity}
\end{figure*}

In Figure \ref{fig:accretedMassDensity}, we plot the total mass density locked into black holes on the left, and its rate of change on the right.  In the left panel, several observational estimates based on integrating luminosity functions, scaling relations, and the X-ray background are over-plotted.  A small black rectangle at $z=0$ shows the estimated SMBH mass density from galaxy scaling relations \citep{Shankar+2009}.  The vertical error bars extending to $z=5$ come from integrating backwards luminosity functions \citep{Hopkins+2007}.  The horizontal error bars around $z=3$ come from stacked X-ray emission \citep{Treister+2009}, as do the short black upper limits  at high redshift \citep{Treister+2013}.  Estimates from the X-ray background for $z>6$ are provided with the black upper limits by \citet{Salvaterra+2012} and a recent reanalysis using a variety of different source spectra by \citet{Cappelluti+2017a}.  From our SAM, we plot the total mass locked into black holes as dashed lines, and the mass derived from luminous accretion as filled regions.  It is the latter set of curves which should satisfy constraints from the X-ray background.  

As can be inferred from the luminosity functions previously discussed, more accretion occurs in the light seed models than in the heavy seed models at high redshifts.  This is because the light seeds are more abundant, and the heavy seeds also get a head start on mass accumulated non-luminously at these early epochs.  On the right panel, this catch-up process results in a second high-redshift bump of AGN activity that peaks around $z=10$, driven entirely by low-luminosity AGN.  By $z=7$, the light seed models have luminously accreted a factor of 2.3 more mass than the heavy seed models, which would approximately correspond to an equivalent difference in X-ray background produced.  Due to the different occupation fractions of the two populations, it is also possible that the backgrounds of these different seeding models may exhibit different clustering properties.  This will be a subject of future work.

There is significant spread in the observational constraints for the $z>6$ data from \citet{Treister+2013}, \citet{Salvaterra+2012}, and \citet{Cappelluti+2017a}.  This is because deriving density constraints from the X-ray background is sensitive to assumptions made about the properties of the source spectra.  A larger amount of accretion can be hidden in the X-ray background by simply reducing the amount of flux assumed to be emitted in the X-ray.  This for instance, would be the case for a population of Compton-thick accretors \citep{Yue+2013}.  Advancing on this front will depend on characterising the spectral energy distributions of $z>6$ AGN with next-generation observatories such as JWST.

\subsection{Gravitational Waves}

SMBH mergers are expected in the hierarchical picture of structure formation, whereby galaxies assemble via the mergers of less-massive ones.  Merging black holes produce gravitational waves, and the first stellar-mass black hole mergers have been detected by the Laser Interferometer Gravitational-Wave Observatory (LIGO) \citep[e.g.,][]{LIGO2016}.  At higher masses, the upcoming Laser Interferometer Space Antenna (LISA) will be sensitive to the intermediate-mass mass black holes that could probe the seed population \citep{LISA2017}.  Gravitational waves have long been recognised as one of the best ways to reconstruct the growth history of black holes from the seeding epoch to present day \citep[see][for a recent comprehensive set of models]{Klein+2016}.  We show here that observations of resolved gravitational wave events will indeed allow us to discriminate between seeding models, irrespective of our assumptions about the steady mode of AGN fuelling.

In order to calculate the distribution of black hole mergers detectable by the future LISA mission, we save the masses, mass ratios, and redshifts of all mergers that occur in our SAM, and then compute their signal-to-noise based on the LISA sensitivity curve.  The signal-to-noise calculation is outlined in \citet{Sesana+2005,Sesana+2007}.  In brief, the signal-to-noise must be integrated in frequency space, since both the strain of the event and the sensitivity of LISA are functions of frequency.  The strain as a function of frequency is calculated based on standard orbital decay calculations, while the sensitivity of LISA is based on the latest payload description document allocation (Neil Cornish, private communication).  We only keep events with a signal-to-noise greater than or equal to 5.  The details of these calculations can be found in Appendix \ref{sec:gravitationalWaveDetails}. 

Unlike the other observables discussed in this work thus far, estimates of the merging black hole population do not depend on their instantaneous accretion state.  On the other hand, they depend on poorly constrained dynamics regarding the formation and shrinking of black hole binaries.  Recall that for our simple dynamical treatment, we assume that BHs merge with only a 10\% probability following a major galaxy merger.  This parameter was added only to limit this growth channel for high-mass SMBHs to prevent them from over-shooting the $M_\bullet-\sigma$ relation at $z=0$.  Since the mass cap for the most massive halos plateaus after $z=6$, BH-BH mergers would otherwise be the dominant mode of growth for the highest mass SMBHs.  This parameterisation with a 10\% probability that a halo merger ultimately results in black hole merger leads to pessimistic numbers for gravitational wave event rates, however.

\begin{table}
\begin{center}
 \begin{tabular}{lll} 
 \hline
 Model & Pessimistic Count & Optimistic Count\\
 \hline
 Light-PL & $118.6 \pm 0.7$ & $979 \pm 2$ \\
 Light-MS & $120.6 \pm 0.6$ & $986 \pm 2$ \\
 Heavy-PL & $19.5 \pm 0.2$  & $194.8 \pm 0.8$ \\
 Heavy-MS & $20.2 \pm 0.2$ & $193.2 \pm 0.7$ \\
 \hline
\end{tabular}
\end{center}
\caption{Total numbers of gravitational wave events predicted by the model, with formal statistical uncertainties.  Pessimistic counts refer to our fiducial runs where $p_\mathrm{merge} = 0.1$, while optimistic counts refer to a different set of runs where $p_\mathrm{merge} = 1$.  Light seeds produce more than an order of magnitude more gravitational wave events than heavy seeds.  Statistical uncertainties from bootstrapping our 20 sets of merger trees are provided.}
\label{tab:gw}
\end{table}

We therefore offer both pessimistic and optimistic event rates for LISA, corresponding to black hole merger probabilities $p_\mathrm{merge}=0.1$ (as in \citetalias{Ricarte&Natarajan2018}) and $p_\mathrm{merge}=1$.  The total number of events produced by each of our models for a four-year mission is provided in Table \ref{tab:gw}.  These range from a few events in the most pessimistic scenarios to hundreds of events in the most optimistic scenarios.  As expected, light seeds produce more than an order of magnitude more events than heavy seeds.  The total number of events detected by LISA will provide a joint constraint on SMBH seeding and the detailed dynamical processes that effect binary black hole mergers.

\begin{figure*}
   \centering
   \includegraphics[width=0.45\textwidth]{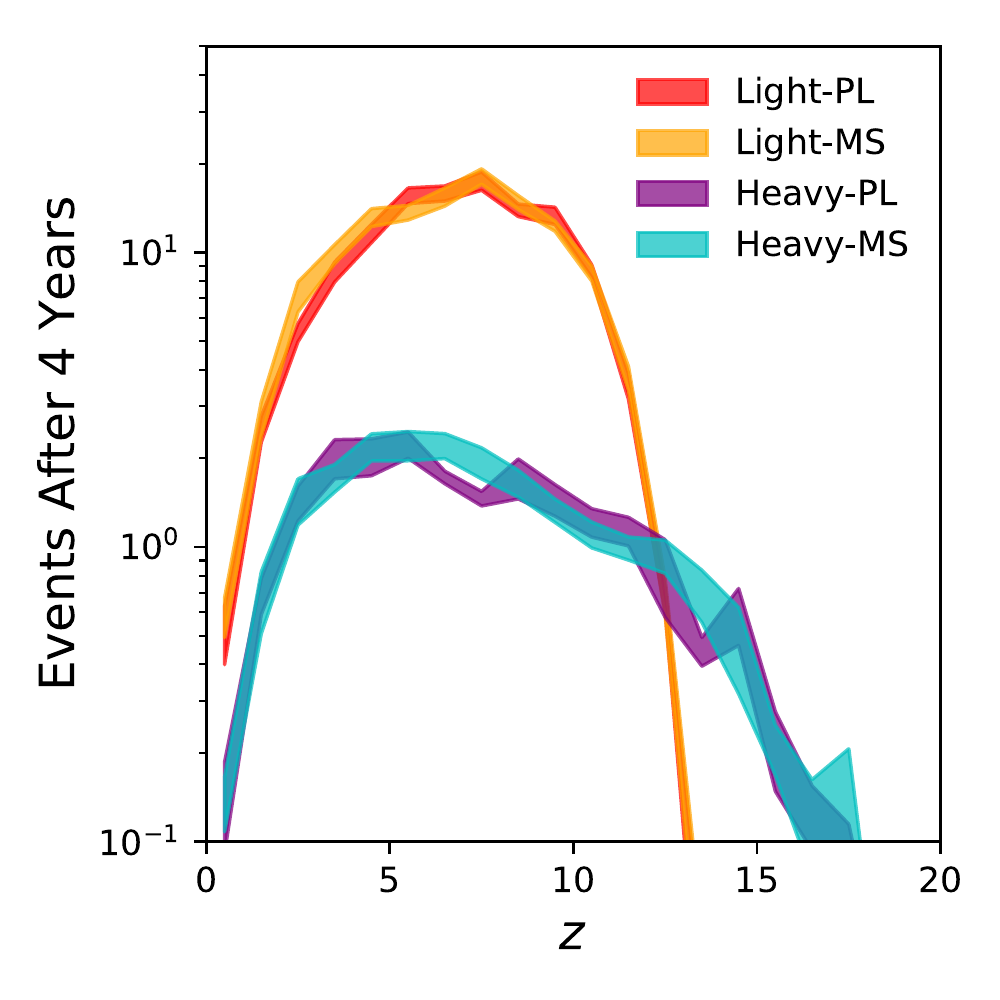}
   \includegraphics[width=0.45\textwidth]{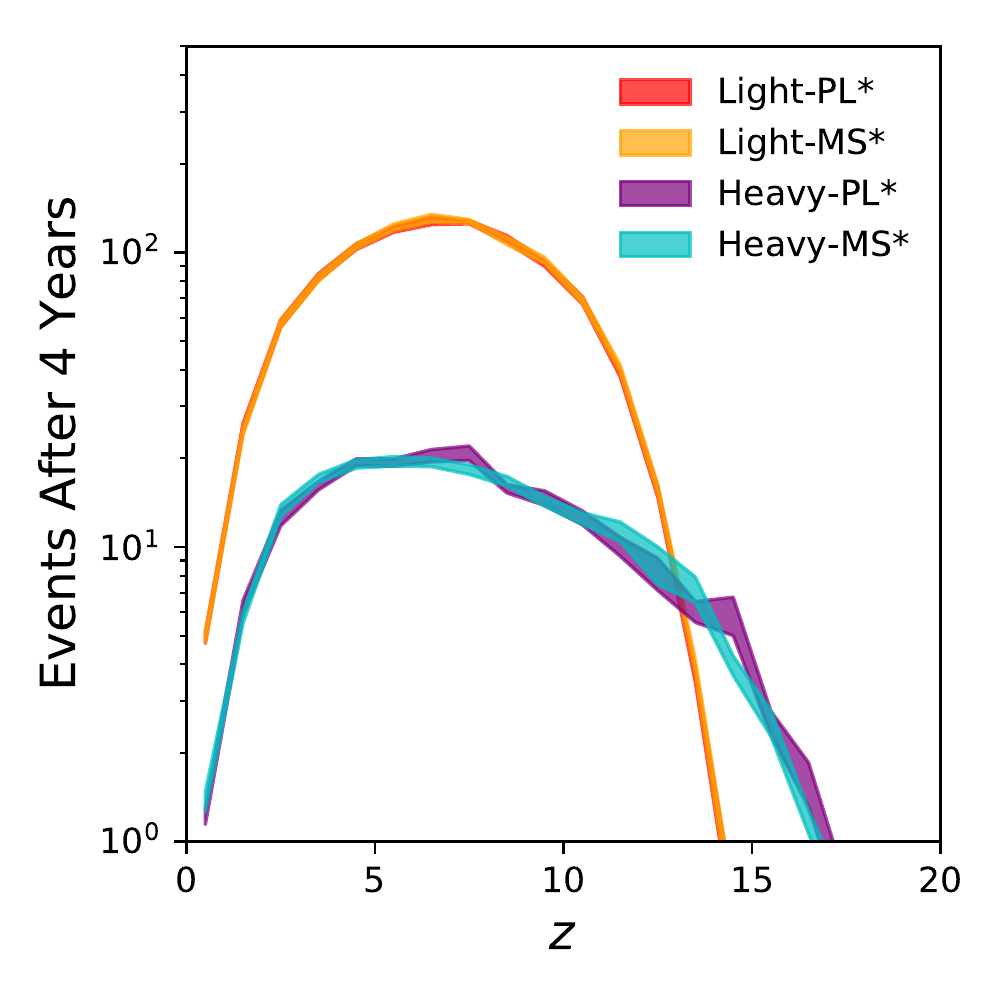} \\
   \includegraphics[width=0.45\textwidth]{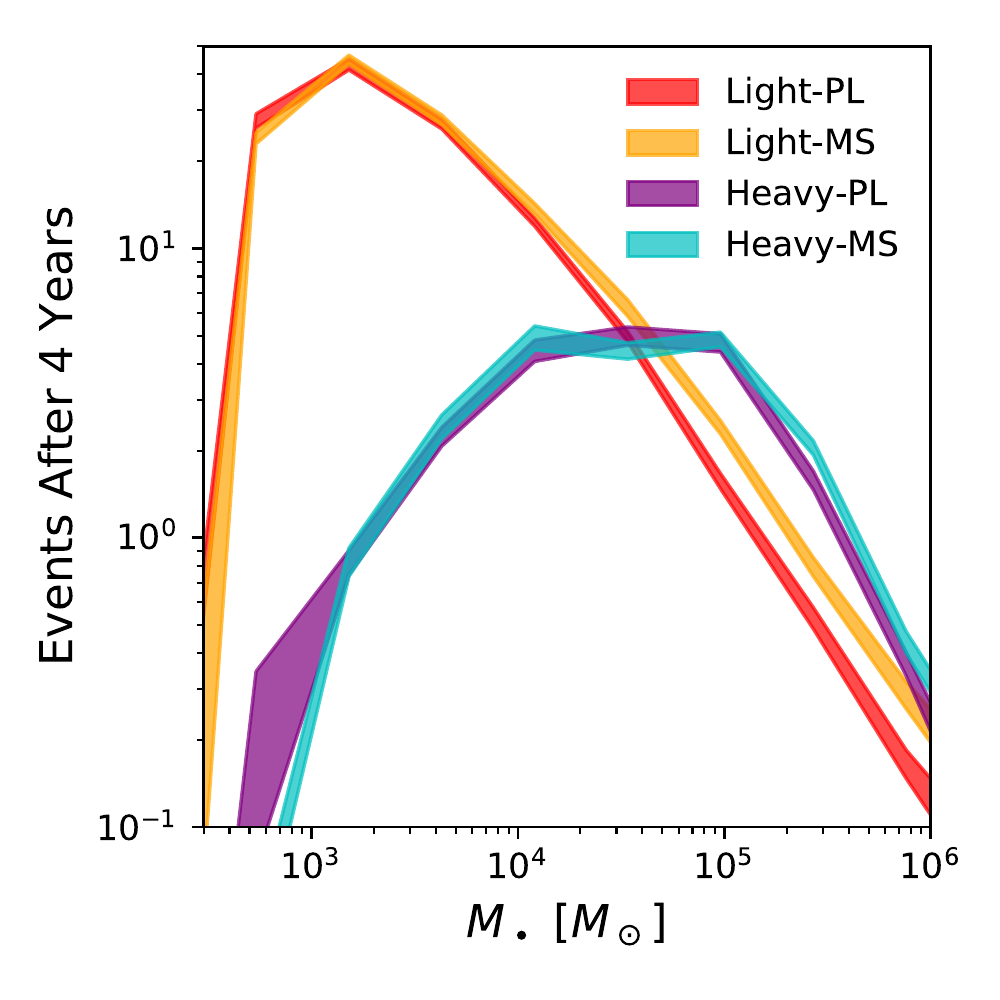}
   \includegraphics[width=0.45\textwidth]{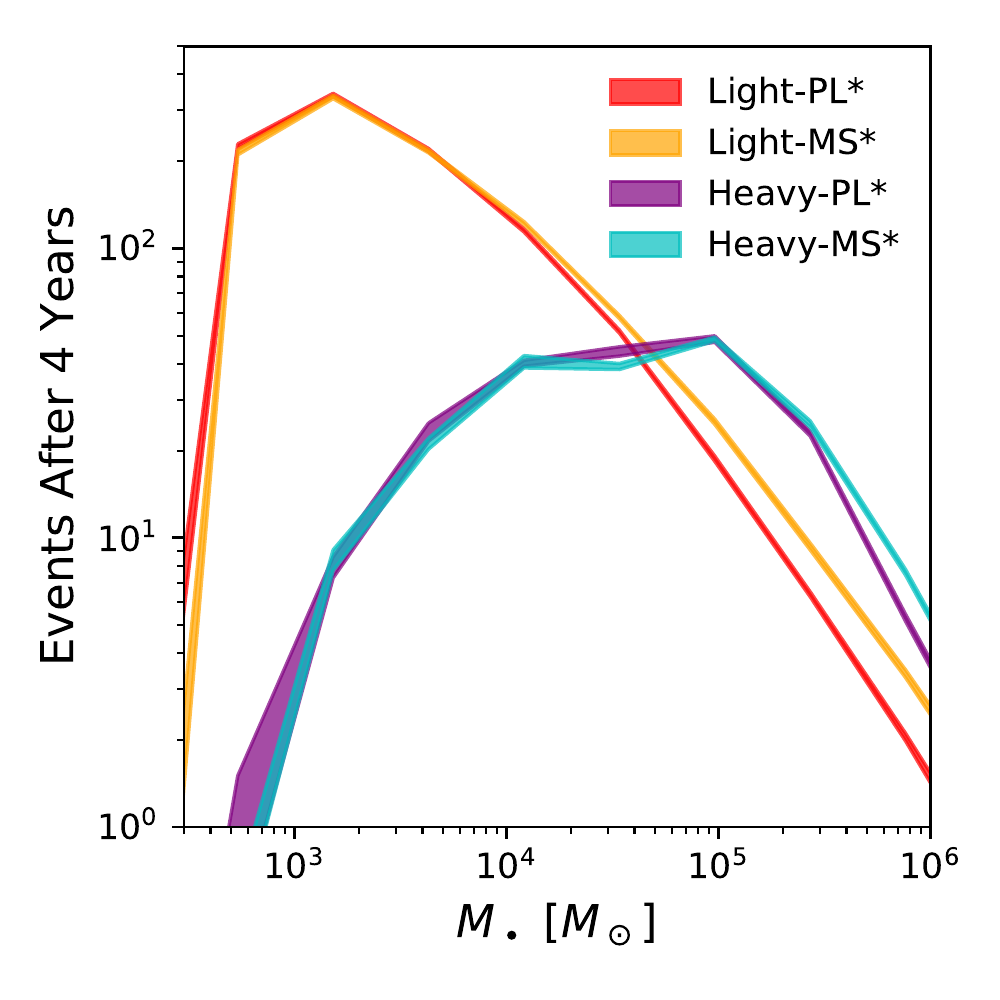} \\
   \caption{Distributions of gravitational wave events detected during a four-year LISA mission, for the pessimistic case on the left and the optimistic case on the right.  {\it Above}:  Distributions as a function of redshift.  The light seeding model produces a more strongly peaked distribution than the heavy seeding model.  {\it Below}:  Distributions as a function of chirp mass.  All models produce a similar number of high-mass events.  However, the heavy models exhibit a drop-off at low-masses that is inherited from the DCBH IMF.}
   \label{fig:lisa_4yr}
\end{figure*}

In addition, we find distinct differences between the two seeding scenarios in the chirp mass and redshift distributions of gravitational wave events.  These are plotted for the pessimistic models in Figure \ref{fig:lisa_4yr} (left column);  for the optimistic case in \ref{fig:lisa_4yr} (right column).  On top, we show the redshift distributions of gravitational wave events in a 3-year time period, while on bottom, we show the chirp mass distributions.  The redshift distributions of light seed mergers are more strongly peaked than those of the heavy seed models.  This, we believe, is due again to the occupation fraction of seeds in low-mass halos.  If a major merger occurs in a low-mass galaxy at high redshift, it is more likely to result in a binary when the occupation fraction is high.  More importantly, the chirp mass distributions of these events contain direct information about the mass distributions of seeds.  In the heavy seed models, a drop-off is expected below $10^5 \ M_\odot$, which is not seen in the light seed models.  Interestingly, the mass distributions above $10^5 \ M_\odot$ are similar for all models, implying that the total numbers of events in Table \ref{tab:gw} for heavy seeds is approximately the total numbers of events for high-mass mergers in even the light models.  Finally, it is encouraging that these rates do not depend on the model for the steady accretion mode. This is because the steady mode does not dominate growth for most black holes in most epochs, except for black holes in the lowest mass hosts at late times.

We remark that the stochastic gravitational wave background, which is dominated by massive, low-redshift sources, does not help distinguish between seeding models.  Nevertheless, the overall normalisation will provide useful information to calibrate event rate calculations.

\section{Discussion}
\label{sec:discussion}

\subsection{Model Limitations}

We have shown that all signatures of SMBH seeds are linked and compounded with our understanding of either accretion or dynamics.  We discuss the limits of each of these aspects of the model.

\subsubsection{Accretion}

In our model, the merger-triggered burst mode dominates SMBH growth until $z=2$, and at lower redshifts the secondary steady mode (either PL or MS) becomes more important.  These are constrained by luminosity functions out to $z=6$.  At $z>6$, the main constraint is that it is necessary to produce $10^9 \ M_\odot$ quasars by $z\sim6$, but there remains some freedom in doing so.  For example, it is possible to assemble the same amount of mass in short, super-Eddington bursts instead of via continuous Eddington limited growth \citep{Pezzulli+2016}.  In addition, the burst mode triggering mechanism itself, major mergers, may not be the primary growth driving mode for the majority of AGN.  Recent studies indicate that while the merger fraction among AGN might be high, not all AGN are mergers \citep{Treister+2012,Hong+2015,Mechtley+2016,Villforth+2017,Weigel+2018}.  In addition, some simulations of the growth of SMBHs in cosmological contexts report that supernova feedback may inhibit the growth of SMBHs in low-mass halos \citep{Dubois+2015,McAlpine+2017,Bower+2017}.  This phenomenon would suppress accretion in our models particularly at high redshift.  

All of these effects may have an impact in reshaping our high-redshift luminosity functions.  Nevertheless, our model does not require any of these changes based on data for $z<6$.  Future refinements of our accretion model will need more guidance from both simulations and multi-wavelength observations.

\subsubsection{Dynamics}

For our gravitational wave calculations, we have provided models with both $p_\mathrm{merge}=0.1$ for a pessimistic case and $p_\mathrm{merge}=1$ for an optimistic case.  The former was used in \citetalias{Ricarte&Natarajan2018} due to the need to limit the BH-BH merger growth channel at high masses.  We comment that the higher value of $p_\mathrm{merge}$ may be accommodated with some trade-offs in the model.  Increasing $p_\mathrm{merge}$ requires that the burst mass cap be lowered for high-mass SMBHs, so that they can still lie on the $M_\bullet-\sigma$ relation at $z=0$.  Since only the high-mass end should be affected, we have experimented with lowering the tilt of the mass cap to $M_\bullet \propto \sigma^4$.  This causes undesirable behaviour at high-redshift, causing the model to produce fewer high-luminosity AGN and more low-luminosity AGN.

Simulations tracking the formation of SMBH pairs in a cosmological context find that $p_\mathrm{merge}$ should increase as a function of host stellar mass and merger mass ratio \citep{Tremmel+2017}.  Implementing more realistic merger probabilities and delay times would be the next step in improving the dynamical treatment of binary black hole mergers. 

\subsection{Other Potential Seeding Signatures}

\subsubsection{The AGN Fraction in Low-mass Galaxies}

The AGN fraction in low-mass galaxies also provides insight into these low-mass BHs and is more easily accessible observationally.  This has been measured to be on the order of $\sim 0.1\%$ for galaxies of stellar mass $\sim 10^9 \ M_\odot$ \citep{Reines+2013,Sartori+2015,Pardo+2016,Aird+2017,Mezcua+2018}.  Due to the intrinsic rarity of AGN in low-mass galaxies, we find that our merger trees do not currently have enough low-mass, low-redshift halos to probe the small AGN fractions that we expect.  In addition, AGN variability on timescales less than the time resolution of our SAM ($\sim 10^7$) \citep{Hickox+2014,Schawinski+2015} may play a significant role in determining AGN fractions, which depend on a luminosity threshold.  Exploring the AGN fraction with a specialised set of merger trees and accretion models would be an interesting avenue for future work.

\subsubsection{Obese Black Hole Galaxies (OBGs)} 

A natural prediction of the heavy seed model is the existence of SMBHs with a higher masses than the stellar mass of their hosts as a transient population at high redshifts.  These are referred to as Obese Black Hole Galaxies (OBGs), and may occur when a halo which formed a DCBH merges with a galaxy.  This is an inversion compared to what we see at $z=0$ locally.  Identifying this transient stage and determining the redshifts at which such sources are detected would be a unique indicator for the DCBH channel \citep{Agarwal+2013}.  Recent work has provided spectral predictions for these sources, and it has been noted that OBGs should be infra-red bright.  Candidates could be found based on their infra-red colours and X-ray emission, and could be analysed with high signal-to-noise with the upcoming JWST mission with both NIRCAM and MIRI \citep{Pacucci+2016,Natarajan+2017}.  Compared to other potentially confusing sources, OBGs are expected to be much redder, with a relatively flat spectral energy distribution out into the mid-infrared.  Detection of even a handful of OBGs would definitely signal that the direct collapse seed formation channel is viable and that it occurs in nature \citep{Natarajan+2017}.

\subsubsection{Clustering in the X-ray Background}

The clustering signal of the X-ray Background might contain information about seeding conditions of high-redshift, unresolved AGN.  In particular, the higher the minimum halo mass which hosts a black hole, the more biased the clustering signal will be.  An excess in the cross-correlation between the X-ray and infrared backgrounds has been reported \citep{Cappelluti+2013,Cappelluti+2017b}, and may be caused by a population of Compton-thick, high-redshift DCBHs \citep{Yue+2013}.  Exploring this hypothesis further with our semi-analytic model will also be the subject of future work. 

\subsection{What about a hybrid model?}

While we have concerned ourselves with distinguishing between either light or heavy seeds, it is most likely that both these channels occur in the universe.  Therefore, we have also experimented with a mixed model in which both light and heavy seeds are present, where the heavy seed mass is taken in case of a conflict.  We report that a mixed model is indistinguishable from a light seed model in all observables except one, the evolution of luminosity functions at the highest redshifts.  Recall from Figure \ref{fig:lum_past6} that at $z \gtrsim 10$, the evolution of the high-luminosity end is governed by the maximum seed mass (plus maximum possible accretion rate), while the evolution of the low-luminosity end is governed by the overall seed abundance.  In the case of a hybrid model, the abundance is high, but there is also a high maximum seed mass.  Qualitatively, the result is a luminosity function which looks like the maximum of both the light and heavy seed curves.  Such a hybrid model has the highest overall luminosity density at these epochs, which may increase the photon budget available for reionisation by AGN.

\section{Conclusion}
\label{sec:conclusion}

The origin of the first seed black holes is currently fiercely debated. While we know that the remnants of the first stars most likely produce light black hole seeds, observations of luminous quasars at the highest redshift suggest that another formation channel that yields massive initial seeds might be required. Using our semi-analytic model to track SMBH assembly, we have examined several observables that may help us distinguish between these light and heavy seeding scenarios. There are two salient features of seeding models that matter:  their abundance, and their maximum mass.  Accretion events based on the host properties and assembly history work to erase the initial conditions, but we still find that unique signatures persist at extremes in both mass and redshift.  These findings are summarised below.

\begin{itemize}
\item {\bf The $M_\bullet-\sigma$ relation}:  While previous studies have claimed that the behaviour at the low-mass end can discriminate between seeding models, we find that the behaviour is shaped primarily by the 
adopted steady accretion mode instead.  The normalisation, slope, and scatter at these low masses is interesting for constraining accretion models, but we do not find that it is very informative for discriminating seeding models.
\item {\bf The Occupation Fraction}:  The occupation fraction is expected to be higher for light seeds than for heavy ones.  We have altered a DCBH seeding parameter to maintain agreement with current constraints from the X-ray emission of low-mass galaxies \citet{Miller+2015}.  We point out, however, that SMBHs don't always grow to their maximal mass at the low-mass end in our model, and therefore may not ``count'' towards the occupation fraction.  We also provide a new metric mass-weighted occupation fractions, which accounts for this effect.  We eagerly anticipate joint constraints from dwarf galaxies and tidal disruption events \citep{Stone&Metzger2016}, for which the modelling uncertainties are very different.
\item {\bf High-redshift Accretion}:  Our light and heavy seeding models produce very different predictions for high-redshift luminosity functions that would be uncovered by Lynx deep field observations.  Light seeds produce many more low-luminosity AGN, due to their increased abundance relative to heavy seeds.  In addition, the high-luminosity end evolves more rapidly with redshift for light seeds than with heavy seeds, since the light seeds become limited by their initial seed mass.  Light seeds would also produce about an order of magnitude more X-ray background emission, allowing some of this information to be accessible without resolving AGN into individual sources.  
\item {\bf Gravitational Wave Events}:  We calculate both redshift- and mass-dependent distributions of predicted gravitational wave events after a three-year LISA mission.  We find that the light seed model predicts a factor of 20 more events, mostly from low-mass coalescences, and that the chirp mass function of gravitational wave events would offer a clean and direct probe of the initial mass function of black hole seeds.
\end{itemize}

These signatures of seeding are all intricately intertwined with our understanding of either accretion processes or dynamics.  Measuring the occupation fraction requires assumptions about accretion, as does interpreting high-redshift luminosity functions and backgrounds.  Meanwhile, the gravitational wave event rate depends on the dynamics that govern the decay of black hole orbits from kpc scales to the gravitational wave regime.  We conclude that the mass distribution of gravitational wave events will provide the cleanest measurement and clearest insight into the seeding mechanism out of all of the observational probes examined here, however, all the probes in combination with be critical to obtain a self-consistent picture of black hole formation and assembly.

\appendix

\section{Computing the Gravitational Wave Event Rate}
\label{sec:gravitationalWaveDetails}

\subsection{The Merger Rate of All Black Holes}

In this Appendix, we outline the procedure using to compute the black hole merger rates from the output of our SAMs. 
During each run of the SAM, we save the primary mass, mass ratio, and redshift of each black hole-black hole merger.  If one has the merger rate per unit comoving volume per unit redshift, $d^2 N/dz dV_c$, the total number of events per observed time interval $dt_\mathrm{obs}$ is given by 

\begin{align}
\frac{dN}{dt_\mathrm{obs}} = \int_0^\infty \frac{d^2 N}{dz dV_c} \frac{dz}{dt_\mathrm{rest}} \frac{dV_c}{dz} \frac{dz}{1+z} .
\end{align}

\noindent where here we have made the substitution $dt_\mathrm{rest} = dt_\mathrm{obs} (1+z)$.  For the list of discrete events, compiled by our SAM, this is instead a summation of detected events $i$, weighted by the abundance of their hosts,

\begin{align}
\frac{dN}{dt_\mathrm{obs}} = \sum_{i} n_\mathrm{host,i} \frac{dz}{dt_\mathrm{rest}}(z_i) \frac{dV_c}{dz}(z_i) \frac{1}{1+z_i} ,
\end{align}

\noindent for all events $i$, where $n_\mathrm{host,i}$ is the comoving number density of the host halo. The remaining derivatives are computed using standard cosmological calculations.

\subsection{Applying the LISA Selection Function} 

In order to compute and predict the number of events detectable by LISA, we need to fold in its sensitivity.
The signal-to-noise of a gravitational wave event is given by

\begin{align}
\frac{S}{N} = \left\{ \int_{f_\mathrm{min}}^{f_\mathrm{ISCO}} d \ln f^\prime \left[ \frac{h_c(f_r^\prime)}{h_\mathrm{rms}(f^\prime)} \right]^2 \right\}^{1/2},
\label{eqn:sn_gw}
\end{align}

\noindent where $h_c(f_r)$ is the event's ``characteristic'' strain at a given frequency, and $h_\mathrm{rms}$ is the sensitivity of LISA as a function of frequency (Neil Cornish, private communication).  To convert the sensitivity curve in units of Hz$^{-1/2}$ into a dimensionless strain, we multiply by $\sqrt{f}$, as appropriate for bursting events \citep{Larson+2000}.

First, let us define the boundaries of this integral.  $f_\mathrm{ISCO}$ is the observed frequency at the innermost stable circular orbit, 

\begin{align}
f_\mathrm{ISCO} = \frac{c^3}{6^{3/2} \pi G}\frac{1}{M_1+M_2}(1+z)^{-1},
\end{align}

\noindent where we have used $f = f_r(1+z)$ to relate observed and rest frequencies.  $f_\mathrm{min}$ is the minimum frequency probed in the lifetime of the observation.  This quantity is calculated by integrating backwards the frequency evolution due to orbital decay over the length of the observation, given by

\begin{align}
\dot{f}_r = \frac{96 \pi^{8/3}G^{5/3}}{5 c^3} \mathcal{M}^{5/3} f_r^{11/3}.
\end{align}

Next, the strain as a function of frequency $h_c(f_r)$ is a function with two parts.  It has a discontinuity at what is defined as $f_\mathrm{break}$, which is the frequency below which the number of cycles, $n$, observed at a given frequency, $f$, satisfies $n > f \tau$, where $\tau$ is the length of the observation.  From orbital decay, the number of cycles as a function of frequency is given by

\begin{align}
n = \frac{5c^5}{96 \pi^{8/3} G^{5/3} \mathcal{M}^{5/3}}f_r^{-5/3},
\end{align}

\noindent so therefore

\begin{align}
f_\mathrm{break}^{8/3} = \frac{5c^5(1+z)^{-5/3}}{96 \pi^{8/3} G^{5/3} \mathcal{M}^{5/3}}.
\end{align}

At rest-frame frequency $f_r$, a gravitational wave event that is a comoving distance $r(z)$ away has a sky- and polarisation-averaged characteristic strain given by 

\begin{align}
h = \frac{8\pi^{2/3}}{10^{1/2}}\frac{G^{5/3}\mathcal{M}^{5/3}}{c^4r(z)}.
\end{align}

\noindent For $f > f_\mathrm{break}$, $h_c(f_r)$ is given by $h_c = h \sqrt{f\tau} \propto f^{7/6}$, while for $f < f_\mathrm{break}$, $h_c = h\sqrt{n} \propto f^{-1/6}$.

These are all of the ingredients required to solve Equation \ref{eqn:sn_gw}.  We keep all events which have a signal-to-noise ratio of at least 5.

\section*{acknowledgements}  This work was supported by NASA Headquarters under the NASA Earth and Space Science Fellowship Program - Grant 80NSSC17K0459.  PN acknowledges support from the NSF TCAN 1332858.
We acknowledge useful discussions during our weekly group meetings with Michael Tremmel, Nico Cappelluti, Fabio Pacucci, Nir Mandelker and Vivienne Baldassare. We also thank Neil Cornish for providing us with the LISA sensitivity function. 

\bibliography{ms}

\begin{thebibliography}{}
\makeatletter
\relax
\def\mn@urlcharsother{\let\do\@makeother \do\$\do\&\do\#\do\^\do\_\do\%\do\~}
\def\mn@doi{\begingroup\mn@urlcharsother \@ifnextchar [ {\mn@doi@}
  {\mn@doi@[]}}
\def\mn@doi@[#1]#2{\def\@tempa{#1}\ifx\@tempa\@empty \href
  {http://dx.doi.org/#2} {doi:#2}\else \href {http://dx.doi.org/#2} {#1}\fi
  \endgroup}
\def\mn@eprint#1#2{\mn@eprint@#1:#2::\@nil}
\def\mn@eprint@arXiv#1{\href {http://arxiv.org/abs/#1} {{\tt arXiv:#1}}}
\def\mn@eprint@dblp#1{\href {http://dblp.uni-trier.de/rec/bibtex/#1.xml}
  {dblp:#1}}
\def\mn@eprint@#1:#2:#3:#4\@nil{\def\@tempa {#1}\def\@tempb {#2}\def\@tempc
  {#3}\ifx \@tempc \@empty \let \@tempc \@tempb \let \@tempb \@tempa \fi \ifx
  \@tempb \@empty \def\@tempb {arXiv}\fi \@ifundefined
  {mn@eprint@\@tempb}{\@tempb:\@tempc}{\expandafter \expandafter \csname
  mn@eprint@\@tempb\endcsname \expandafter{\@tempc}}}

\bibitem[\protect\citeauthoryear{Abbott et~al.,}{Abbott
  et~al.}{2016}]{LIGO2016}
Abbott B.~P.,  et~al., 2016, \mn@doi [Phys. Rev. Lett.]
  {10.1103/PhysRevLett.116.061102}, 116, 061102

\bibitem[\protect\citeauthoryear{{Agarwal}, {Davis}, {Khochfar}, {Natarajan}
  \& {Dunlop}}{{Agarwal} et~al.}{2013}]{Agarwal+2013}
{Agarwal} B.,  {Davis} A.~J.,  {Khochfar} S.,  {Natarajan} P.,   {Dunlop}
  J.~S.,  2013, \mn@doi [\mnras] {10.1093/mnras/stt696}, \href
  {http://adsabs.harvard.edu/abs/2013MNRAS.432.3438A} {432, 3438}

\bibitem[\protect\citeauthoryear{{Agarwal}, {Smith}, {Glover}, {Natarajan}  \&
  {Khochfar}}{{Agarwal} et~al.}{2016}]{Agarwal+2016}
{Agarwal} B.,  {Smith} B.,  {Glover} S.,  {Natarajan} P.,   {Khochfar} S.,
  2016, \mn@doi [\mnras] {10.1093/mnras/stw929}, \href
  {http://adsabs.harvard.edu/abs/2016MNRAS.459.4209A} {459, 4209}

\bibitem[\protect\citeauthoryear{{Aird}, {Coil}  \& {Georgakakis}}{{Aird}
  et~al.}{2017}]{Aird+2017}
{Aird} J.,  {Coil} A.~L.,   {Georgakakis} A.,  2017, \mn@doi [\mnras]
  {10.1093/mnras/stw2932}, \href
  {http://adsabs.harvard.edu/abs/2017MNRAS.465.3390A} {465, 3390}

\bibitem[\protect\citeauthoryear{{Alexander} \& {Natarajan}}{{Alexander} \&
  {Natarajan}}{2014}]{Alexander&Natarajan2014}
{Alexander} T.,  {Natarajan} P.,  2014, \mn@doi [Science]
  {10.1126/science.1251053}, \href
  {http://adsabs.harvard.edu/abs/2014Sci...345.1330A} {345, 1330}

\bibitem[\protect\citeauthoryear{{Amaro-Seoane} et~al.,}{{Amaro-Seoane}
  et~al.}{2017}]{LISA2017}
{Amaro-Seoane} P.,  et~al., 2017, preprint, \href
  {http://adsabs.harvard.edu/abs/2017arXiv170200786A} {} (\mn@eprint {arXiv}
  {1702.00786})

\bibitem[\protect\citeauthoryear{{Ba{\~n}ados} et~al.,}{{Ba{\~n}ados}
  et~al.}{2018}]{Banados+2018}
{Ba{\~n}ados} E.,  et~al., 2018, \mn@doi [\nat] {10.1038/nature25180}, \href
  {http://adsabs.harvard.edu/abs/2018Natur.553..473B} {553, 473}

\bibitem[\protect\citeauthoryear{{Baldassare}, {Reines}, {Gallo}  \&
  {Greene}}{{Baldassare} et~al.}{2015}]{Baldassare+2015}
{Baldassare} V.~F.,  {Reines} A.~E.,  {Gallo} E.,   {Greene} J.~E.,  2015,
  \mn@doi [\apjl] {10.1088/2041-8205/809/1/L14}, \href
  {http://adsabs.harvard.edu/abs/2015ApJ...809L..14B} {809, L14}

\bibitem[\protect\citeauthoryear{{Baldassare} et~al.,}{{Baldassare}
  et~al.}{2016}]{Baldassare+2016}
{Baldassare} V.~F.,  et~al., 2016, \mn@doi [\apj] {10.3847/0004-637X/829/1/57},
  \href {http://adsabs.harvard.edu/abs/2016ApJ...829...57B} {829, 57}

\bibitem[\protect\citeauthoryear{{Barth}, {Ho}, {Rutledge}  \&
  {Sargent}}{{Barth} et~al.}{2004}]{Barth+2004}
{Barth} A.~J.,  {Ho} L.~C.,  {Rutledge} R.~E.,   {Sargent} W.~L.~W.,  2004,
  \mn@doi [\apj] {10.1086/383302}, \href
  {http://adsabs.harvard.edu/abs/2004ApJ...607...90B} {607, 90}

\bibitem[\protect\citeauthoryear{{Begelman}, {Volonteri}  \& {Rees}}{{Begelman}
  et~al.}{2006}]{Begelman+2006}
{Begelman} M.~C.,  {Volonteri} M.,   {Rees} M.~J.,  2006, \mn@doi [\mnras]
  {10.1111/j.1365-2966.2006.10467.x}, \href
  {http://adsabs.harvard.edu/abs/2006MNRAS.370..289B} {370, 289}

\bibitem[\protect\citeauthoryear{{Ben-Ami}, {Vikhlinin}  \& {Loeb}}{{Ben-Ami}
  et~al.}{2018}]{Ben-Ami+2018}
{Ben-Ami} S.,  {Vikhlinin} A.,   {Loeb} A.,  2018, \mn@doi [\apj]
  {10.3847/1538-4357/aaa6d0}, \href
  {http://adsabs.harvard.edu/abs/2018ApJ...854....4B} {854, 4}

\bibitem[\protect\citeauthoryear{{Bird}, {Cholis}, {Mu{\~n}oz},
  {Ali-Ha{\"i}moud}, {Kamionkowski}, {Kovetz}, {Raccanelli}  \& {Riess}}{{Bird}
  et~al.}{2016}]{Bird+2016}
{Bird} S.,  {Cholis} I.,  {Mu{\~n}oz} J.~B.,  {Ali-Ha{\"i}moud} Y.,
  {Kamionkowski} M.,  {Kovetz} E.~D.,  {Raccanelli} A.,   {Riess} A.~G.,  2016,
  \mn@doi [Physical Review Letters] {10.1103/PhysRevLett.116.201301}, \href
  {http://adsabs.harvard.edu/abs/2016PhRvL.116t1301B} {116, 201301}

\bibitem[\protect\citeauthoryear{{Bower}, {Schaye}, {Frenk}, {Theuns},
  {Schaller}, {Crain}  \& {McAlpine}}{{Bower} et~al.}{2017}]{Bower+2017}
{Bower} R.~G.,  {Schaye} J.,  {Frenk} C.~S.,  {Theuns} T.,  {Schaller} M.,
  {Crain} R.~A.,   {McAlpine} S.,  2017, \mn@doi [\mnras]
  {10.1093/mnras/stw2735}, \href
  {http://adsabs.harvard.edu/abs/2017MNRAS.465...32B} {465, 32}

\bibitem[\protect\citeauthoryear{{Boylan-Kolchin}, {Ma}  \&
  {Quataert}}{{Boylan-Kolchin} et~al.}{2008}]{Boylan-Kolchin+2008}
{Boylan-Kolchin} M.,  {Ma} C.-P.,   {Quataert} E.,  2008, \mn@doi [\mnras]
  {10.1111/j.1365-2966.2007.12530.x}, \href
  {http://adsabs.harvard.edu/abs/2008MNRAS.383...93B} {383, 93}

\bibitem[\protect\citeauthoryear{{Bromm} \& {Loeb}}{{Bromm} \&
  {Loeb}}{2003}]{Bromm&Loeb2003}
{Bromm} V.,  {Loeb} A.,  2003, \mn@doi [\apj] {10.1086/377529}, \href
  {http://adsabs.harvard.edu/abs/2003ApJ...596...34B} {596, 34}

\bibitem[\protect\citeauthoryear{{Bromm}, {Coppi}  \& {Larson}}{{Bromm}
  et~al.}{2002}]{Bromm+2002}
{Bromm} V.,  {Coppi} P.~S.,   {Larson} R.~B.,  2002, \mn@doi [\apj]
  {10.1086/323947}, \href {http://adsabs.harvard.edu/abs/2002ApJ...564...23B}
  {564, 23}

\bibitem[\protect\citeauthoryear{{Cappelluti} et~al.,}{{Cappelluti}
  et~al.}{2013}]{Cappelluti+2013}
{Cappelluti} N.,  et~al., 2013, \mn@doi [\apj] {10.1088/0004-637X/769/1/68},
  \href {http://adsabs.harvard.edu/abs/2013ApJ...769...68C} {769, 68}

\bibitem[\protect\citeauthoryear{{Cappelluti} et~al.,}{{Cappelluti}
  et~al.}{2017a}]{Cappelluti+2017a}
{Cappelluti} N.,  et~al., 2017a, \mn@doi [\apj] {10.3847/1538-4357/aa5ea4},
  \href {http://adsabs.harvard.edu/abs/2017ApJ...837...19C} {837, 19}

\bibitem[\protect\citeauthoryear{{Cappelluti} et~al.,}{{Cappelluti}
  et~al.}{2017b}]{Cappelluti+2017b}
{Cappelluti} N.,  et~al., 2017b, \mn@doi [\apjl] {10.3847/2041-8213/aa8acd},
  \href {http://adsabs.harvard.edu/abs/2017ApJ...847L..11C} {847, L11}

\bibitem[\protect\citeauthoryear{{Choksi}, {Behroozi}, {Volonteri},
  {Schneider}, {Ma}, {Silk}  \& {Moster}}{{Choksi} et~al.}{2017}]{Choksi+2017}
{Choksi} N.,  {Behroozi} P.,  {Volonteri} M.,  {Schneider} R.,  {Ma} C.-P.,
  {Silk} J.,   {Moster} B.,  2017, \mn@doi [\mnras] {10.1093/mnras/stx2113},
  \href {https://ui.adsabs.harvard.edu/#abs/2017MNRAS.472.1526C} {472, 1526}

\bibitem[\protect\citeauthoryear{{Chon}, {Hirano}, {Hosokawa}  \&
  {Yoshida}}{{Chon} et~al.}{2016}]{Chon+2016}
{Chon} S.,  {Hirano} S.,  {Hosokawa} T.,   {Yoshida} N.,  2016, \mn@doi [\apj]
  {10.3847/0004-637X/832/2/134}, \href
  {http://adsabs.harvard.edu/abs/2016ApJ...832..134C} {832, 134}

\bibitem[\protect\citeauthoryear{{Clark}, {Glover}, {Klessen}  \&
  {Bromm}}{{Clark} et~al.}{2011}]{Clark+2011}
{Clark} P.~C.,  {Glover} S.~C.~O.,  {Klessen} R.~S.,   {Bromm} V.,  2011,
  \mn@doi [\apj] {10.1088/0004-637X/727/2/110}, \href
  {http://adsabs.harvard.edu/abs/2011ApJ...727..110C} {727, 110}

\bibitem[\protect\citeauthoryear{{Colpi}}{{Colpi}}{2014}]{Colpi2014}
{Colpi} M.,  2014, \mn@doi [\ssr] {10.1007/s11214-014-0067-1}, \href
  {https://ui.adsabs.harvard.edu/#abs/2014SSRv..183..189C} {183, 189}

\bibitem[\protect\citeauthoryear{{Cowie}, {Barger}  \& {Hasinger}}{{Cowie}
  et~al.}{2012}]{Cowie+2012}
{Cowie} L.~L.,  {Barger} A.~J.,   {Hasinger} G.,  2012, \mn@doi [\apj]
  {10.1088/0004-637X/748/1/50}, \href
  {http://adsabs.harvard.edu/abs/2012ApJ...748...50C} {748, 50}

\bibitem[\protect\citeauthoryear{{Desroches} \& {Ho}}{{Desroches} \&
  {Ho}}{2009}]{Desroches&Ho2009}
{Desroches} L.-B.,  {Ho} L.~C.,  2009, \mn@doi [\apj]
  {10.1088/0004-637X/690/1/267}, \href
  {http://adsabs.harvard.edu/abs/2009ApJ...690..267D} {690, 267}

\bibitem[\protect\citeauthoryear{{Devecchi} \& {Volonteri}}{{Devecchi} \&
  {Volonteri}}{2009}]{Devecchi&Volonteri2009}
{Devecchi} B.,  {Volonteri} M.,  2009, \mn@doi [\apj]
  {10.1088/0004-637X/694/1/302}, \href
  {http://adsabs.harvard.edu/abs/2009ApJ...694..302D} {694, 302}

\bibitem[\protect\citeauthoryear{{Dubois}, {Volonteri}, {Silk}, {Devriendt},
  {Slyz}  \& {Teyssier}}{{Dubois} et~al.}{2015}]{Dubois+2015}
{Dubois} Y.,  {Volonteri} M.,  {Silk} J.,  {Devriendt} J.,  {Slyz} A.,
  {Teyssier} R.,  2015, \mn@doi [\mnras] {10.1093/mnras/stv1416}, \href
  {http://adsabs.harvard.edu/abs/2015MNRAS.452.1502D} {452, 1502}

\bibitem[\protect\citeauthoryear{{Fan} et~al.,}{{Fan} et~al.}{2003}]{Fan+2003}
{Fan} X.,  et~al., 2003, \mn@doi [\aj] {10.1086/368246}, \href
  {http://adsabs.harvard.edu/abs/2003AJ....125.1649F} {125, 1649}

\bibitem[\protect\citeauthoryear{{Ferrara}, {Salvadori}, {Yue}  \&
  {Schleicher}}{{Ferrara} et~al.}{2014}]{Ferrara+2014}
{Ferrara} A.,  {Salvadori} S.,  {Yue} B.,   {Schleicher} D.,  2014, \mn@doi
  [\mnras] {10.1093/mnras/stu1280}, \href
  {http://adsabs.harvard.edu/abs/2014MNRAS.443.2410F} {443, 2410}

\bibitem[\protect\citeauthoryear{{Ferrarese} \& {Merritt}}{{Ferrarese} \&
  {Merritt}}{2000}]{Ferrarese&Merritt2000}
{Ferrarese} L.,  {Merritt} D.,  2000, \mn@doi [\apjl] {10.1086/312838}, \href
  {http://adsabs.harvard.edu/abs/2000ApJ...539L...9F} {539, L9}

\bibitem[\protect\citeauthoryear{{Gallo}, {Treu}, {Marshall}, {Woo}, {Leipski}
  \& {Antonucci}}{{Gallo} et~al.}{2010}]{Gallo+2010}
{Gallo} E.,  {Treu} T.,  {Marshall} P.~J.,  {Woo} J.-H.,  {Leipski} C.,
  {Antonucci} R.,  2010, \mn@doi [\apj] {10.1088/0004-637X/714/1/25}, \href
  {http://adsabs.harvard.edu/abs/2010ApJ...714...25G} {714, 25}

\bibitem[\protect\citeauthoryear{{Greene}}{{Greene}}{2012}]{Greene2012}
{Greene} J.~E.,  2012, \mn@doi [Nature Communications] {10.1038/ncomms2314},
  \href {http://adsabs.harvard.edu/abs/2012NatCo...3E1304G} {3, 1304}

\bibitem[\protect\citeauthoryear{{Greif}, {Bromm}, {Clark}, {Glover}, {Smith},
  {Klessen}, {Yoshida}  \& {Springel}}{{Greif} et~al.}{2012}]{Greif+2012}
{Greif} T.~H.,  {Bromm} V.,  {Clark} P.~C.,  {Glover} S.~C.~O.,  {Smith} R.~J.,
   {Klessen} R.~S.,  {Yoshida} N.,   {Springel} V.,  2012, \mn@doi [\mnras]
  {10.1111/j.1365-2966.2012.21212.x}, \href
  {http://adsabs.harvard.edu/abs/2012MNRAS.424..399G} {424, 399}

\bibitem[\protect\citeauthoryear{{Haiman}}{{Haiman}}{2013}]{Haiman2013}
{Haiman} Z.,  2013, in {Wiklind} T.,  {Mobasher} B.,   {Bromm} V.,  eds,
  Astrophysics and Space Science Library Vol. 396, The First Galaxies. p.~293
  (\mn@eprint {arXiv} {1203.6075}), \mn@doi{10.1007/978-3-642-32362-1_6}

\bibitem[\protect\citeauthoryear{{Hickox}, {Mullaney}, {Alexander}, {Chen},
  {Civano}, {Goulding}  \& {Hainline}}{{Hickox} et~al.}{2014}]{Hickox+2014}
{Hickox} R.~C.,  {Mullaney} J.~R.,  {Alexander} D.~M.,  {Chen} C.-T.~J.,
  {Civano} F.~M.,  {Goulding} A.~D.,   {Hainline} K.~N.,  2014, \mn@doi [\apj]
  {10.1088/0004-637X/782/1/9}, \href
  {http://adsabs.harvard.edu/abs/2014ApJ...782....9H} {782, 9}

\bibitem[\protect\citeauthoryear{{Hirano}, {Hosokawa}, {Yoshida}, {Umeda},
  {Omukai}, {Chiaki}  \& {Yorke}}{{Hirano} et~al.}{2014}]{Hirano+2014}
{Hirano} S.,  {Hosokawa} T.,  {Yoshida} N.,  {Umeda} H.,  {Omukai} K.,
  {Chiaki} G.,   {Yorke} H.~W.,  2014, \mn@doi [\apj]
  {10.1088/0004-637X/781/2/60}, \href
  {http://adsabs.harvard.edu/abs/2014ApJ...781...60H} {781, 60}

\bibitem[\protect\citeauthoryear{{Hong}, {Im}, {Kim}  \& {Ho}}{{Hong}
  et~al.}{2015}]{Hong+2015}
{Hong} J.,  {Im} M.,  {Kim} M.,   {Ho} L.~C.,  2015, \mn@doi [\apj]
  {10.1088/0004-637X/804/1/34}, \href
  {http://adsabs.harvard.edu/abs/2015ApJ...804...34H} {804, 34}

\bibitem[\protect\citeauthoryear{{Hopkins}, {Richards}  \&
  {Hernquist}}{{Hopkins} et~al.}{2007}]{Hopkins+2007}
{Hopkins} P.~F.,  {Richards} G.~T.,   {Hernquist} L.,  2007, \mn@doi [\apj]
  {10.1086/509629}, \href {http://adsabs.harvard.edu/abs/2007ApJ...654..731H}
  {654, 731}

\bibitem[\protect\citeauthoryear{{Huertas-Company} et~al.,}{{Huertas-Company}
  et~al.}{2013}]{Huertas-Company+2013}
{Huertas-Company} M.,  et~al., 2013, \mn@doi [\mnras] {10.1093/mnras/sts150},
  \href {http://adsabs.harvard.edu/abs/2013MNRAS.428.1715H} {428, 1715}

\bibitem[\protect\citeauthoryear{{Inayoshi}, {Haiman}  \&
  {Ostriker}}{{Inayoshi} et~al.}{2016}]{Inayoshi+2016}
{Inayoshi} K.,  {Haiman} Z.,   {Ostriker} J.~P.,  2016, \mn@doi [\mnras]
  {10.1093/mnras/stw836}, \href
  {http://adsabs.harvard.edu/abs/2016MNRAS.459.3738I} {459, 3738}

\bibitem[\protect\citeauthoryear{{Jahnke} \& {Macci{\`o}}}{{Jahnke} \&
  {Macci{\`o}}}{2011}]{Jahnke&Maccio2011}
{Jahnke} K.,  {Macci{\`o}} A.~V.,  2011, \mn@doi [\apj]
  {10.1088/0004-637X/734/2/92}, \href
  {http://adsabs.harvard.edu/abs/2011ApJ...734...92J} {734, 92}

\bibitem[\protect\citeauthoryear{{King}}{{King}}{2003}]{King2003}
{King} A.,  2003, \mn@doi [\apjl] {10.1086/379143}, \href
  {http://adsabs.harvard.edu/abs/2003ApJ...596L..27K} {596, L27}

\bibitem[\protect\citeauthoryear{{King} \& {Pounds}}{{King} \&
  {Pounds}}{2015}]{King&Pounds2015}
{King} A.,  {Pounds} K.,  2015, \mn@doi [\araa]
  {10.1146/annurev-astro-082214-122316}, \href
  {http://adsabs.harvard.edu/abs/2015ARA%26A..53..115K} {53, 115}

\bibitem[\protect\citeauthoryear{{Klein} et~al.,}{{Klein}
  et~al.}{2016}]{Klein+2016}
{Klein} A.,  et~al., 2016, \mn@doi [\prd] {10.1103/PhysRevD.93.024003}, \href
  {https://ui.adsabs.harvard.edu/#abs/2016PhRvD..93b4003K} {93, 024003}

\bibitem[\protect\citeauthoryear{{Kormendy} \& {Ho}}{{Kormendy} \&
  {Ho}}{2013}]{Kormendy&Ho2013}
{Kormendy} J.,  {Ho} L.~C.,  2013, \mn@doi [\araa]
  {10.1146/annurev-astro-082708-101811}, \href
  {http://adsabs.harvard.edu/abs/2013ARA%26A..51..511K} {51, 511}

\bibitem[\protect\citeauthoryear{{Larson}, {Hiscock}  \& {Hellings}}{{Larson}
  et~al.}{2000}]{Larson+2000}
{Larson} S.~L.,  {Hiscock} W.~A.,   {Hellings} R.~W.,  2000, \mn@doi [\prd]
  {10.1103/PhysRevD.62.062001}, \href
  {http://adsabs.harvard.edu/abs/2000PhRvD..62f2001L} {62, 062001}

\bibitem[\protect\citeauthoryear{{Latif}, {Schleicher}, {Schmidt}  \&
  {Niemeyer}}{{Latif} et~al.}{2013}]{Latif+2013}
{Latif} M.~A.,  {Schleicher} D.~R.~G.,  {Schmidt} W.,   {Niemeyer} J.,  2013,
  \mn@doi [\apjl] {10.1088/2041-8205/772/1/L3}, \href
  {http://adsabs.harvard.edu/abs/2013ApJ...772L...3L} {772, L3}

\bibitem[\protect\citeauthoryear{{Lodato} \& {Natarajan}}{{Lodato} \&
  {Natarajan}}{2006}]{Lodato&Natarajan2006}
{Lodato} G.,  {Natarajan} P.,  2006, \mn@doi [\mnras]
  {10.1111/j.1365-2966.2006.10801.x}, \href
  {http://adsabs.harvard.edu/abs/2006MNRAS.371.1813L} {371, 1813}

\bibitem[\protect\citeauthoryear{{Lodato} \& {Natarajan}}{{Lodato} \&
  {Natarajan}}{2007}]{Lodato&Natarajan2007}
{Lodato} G.,  {Natarajan} P.,  2007, \mn@doi [\mnras]
  {10.1111/j.1745-3933.2007.00304.x}, \href
  {http://adsabs.harvard.edu/abs/2007MNRAS.377L..64L} {377, L64}

\bibitem[\protect\citeauthoryear{{Lousto}, {Zlochower}, {Dotti}  \&
  {Volonteri}}{{Lousto} et~al.}{2012}]{Lousto+2012}
{Lousto} C.~O.,  {Zlochower} Y.,  {Dotti} M.,   {Volonteri} M.,  2012, \mn@doi
  [\prd] {10.1103/PhysRevD.85.084015}, \href
  {http://adsabs.harvard.edu/abs/2012PhRvD..85h4015L} {85, 084015}

\bibitem[\protect\citeauthoryear{{McAlpine}, {Bower}, {Harrison}, {Crain},
  {Schaller}, {Schaye}  \& {Theuns}}{{McAlpine} et~al.}{2017}]{McAlpine+2017}
{McAlpine} S.,  {Bower} R.~G.,  {Harrison} C.~M.,  {Crain} R.~A.,  {Schaller}
  M.,  {Schaye} J.,   {Theuns} T.,  2017, \mn@doi [\mnras]
  {10.1093/mnras/stx658}, \href
  {http://adsabs.harvard.edu/abs/2017MNRAS.468.3395M} {468, 3395}

\bibitem[\protect\citeauthoryear{{Mechtley} et~al.,}{{Mechtley}
  et~al.}{2016}]{Mechtley+2016}
{Mechtley} M.,  et~al., 2016, \mn@doi [\apj] {10.3847/0004-637X/830/2/156},
  \href {http://adsabs.harvard.edu/abs/2016ApJ...830..156M} {830, 156}

\bibitem[\protect\citeauthoryear{{Mezcua}, {Civano}, {Marchesi}, {Suh},
  {Fabbiano}  \& {Volonteri}}{{Mezcua} et~al.}{2018}]{Mezcua+2018}
{Mezcua} M.,  {Civano} F.,  {Marchesi} S.,  {Suh} H.,  {Fabbiano} G.,
  {Volonteri} M.,  2018, preprint, \href
  {http://adsabs.harvard.edu/abs/2018arXiv180201567M} {} (\mn@eprint {arXiv}
  {1802.01567})

\bibitem[\protect\citeauthoryear{{Miller}, {Gallo}, {Greene}, {Kelly}, {Treu},
  {Woo}  \& {Baldassare}}{{Miller} et~al.}{2015}]{Miller+2015}
{Miller} B.~P.,  {Gallo} E.,  {Greene} J.~E.,  {Kelly} B.~C.,  {Treu} T.,
  {Woo} J.-H.,   {Baldassare} V.,  2015, \mn@doi [\apj]
  {10.1088/0004-637X/799/1/98}, \href
  {http://adsabs.harvard.edu/abs/2015ApJ...799...98M} {799, 98}

\bibitem[\protect\citeauthoryear{{Mortlock} et~al.,}{{Mortlock}
  et~al.}{2011}]{Mortlock+2011}
{Mortlock} D.~J.,  et~al., 2011, \mn@doi [\nat] {10.1038/nature10159}, \href
  {http://adsabs.harvard.edu/abs/2011Natur.474..616M} {474, 616}

\bibitem[\protect\citeauthoryear{{Mosleh}, {Williams}  \& {Franx}}{{Mosleh}
  et~al.}{2013}]{Mosleh+2013}
{Mosleh} M.,  {Williams} R.~J.,   {Franx} M.,  2013, \mn@doi [\apj]
  {10.1088/0004-637X/777/2/117}, \href
  {http://adsabs.harvard.edu/abs/2013ApJ...777..117M} {777, 117}

\bibitem[\protect\citeauthoryear{{Moster}, {Naab}  \& {White}}{{Moster}
  et~al.}{2013}]{Moster+2013}
{Moster} B.~P.,  {Naab} T.,   {White} S.~D.~M.,  2013, \mn@doi [\mnras]
  {10.1093/mnras/sts261}, \href
  {http://adsabs.harvard.edu/abs/2013MNRAS.428.3121M} {428, 3121}

\bibitem[\protect\citeauthoryear{{Mullaney} et~al.,}{{Mullaney}
  et~al.}{2012}]{Mullaney+2012}
{Mullaney} J.~R.,  et~al., 2012, \mn@doi [\apjl] {10.1088/2041-8205/753/2/L30},
  \href {http://adsabs.harvard.edu/abs/2012ApJ...753L..30M} {753, L30}

\bibitem[\protect\citeauthoryear{{Natarajan}}{{Natarajan}}{2014}]{Natarajan2014}
{Natarajan} P.,  2014, \mn@doi [General Relativity and Gravitation]
  {10.1007/s10714-014-1702-6}, \href
  {http://adsabs.harvard.edu/abs/2014GReGr..46.1702N} {46, 1702}

\bibitem[\protect\citeauthoryear{{Natarajan} \& {Treister}}{{Natarajan} \&
  {Treister}}{2009}]{Natarajan&Treister2009}
{Natarajan} P.,  {Treister} E.,  2009, \mn@doi [\mnras]
  {10.1111/j.1365-2966.2008.13864.x}, \href
  {http://adsabs.harvard.edu/abs/2009MNRAS.393..838N} {393, 838}

\bibitem[\protect\citeauthoryear{{Natarajan}, {Pacucci}, {Ferrara}, {Agarwal},
  {Ricarte}, {Zackrisson}  \& {Cappelluti}}{{Natarajan}
  et~al.}{2017}]{Natarajan+2017}
{Natarajan} P.,  {Pacucci} F.,  {Ferrara} A.,  {Agarwal} B.,  {Ricarte} A.,
  {Zackrisson} E.,   {Cappelluti} N.,  2017, \mn@doi [\apj]
  {10.3847/1538-4357/aa6330}, \href
  {http://adsabs.harvard.edu/abs/2017ApJ...838..117N} {838, 117}

\bibitem[\protect\citeauthoryear{{Pacucci}, {Ferrara}, {Grazian}, {Fiore},
  {Giallongo}  \& {Puccetti}}{{Pacucci} et~al.}{2016}]{Pacucci+2016}
{Pacucci} F.,  {Ferrara} A.,  {Grazian} A.,  {Fiore} F.,  {Giallongo} E.,
  {Puccetti} S.,  2016, \mn@doi [\mnras] {10.1093/mnras/stw725}, \href
  {http://adsabs.harvard.edu/abs/2016MNRAS.459.1432P} {459, 1432}

\bibitem[\protect\citeauthoryear{{Pacucci}, {Natarajan}, {Volonteri},
  {Cappelluti}  \& {Urry}}{{Pacucci} et~al.}{2017}]{Pacucci+2017}
{Pacucci} F.,  {Natarajan} P.,  {Volonteri} M.,  {Cappelluti} N.,   {Urry}
  C.~M.,  2017, \mn@doi [\apjl] {10.3847/2041-8213/aa9aea}, \href
  {http://adsabs.harvard.edu/abs/2017ApJ...850L..42P} {850, L42}

\bibitem[\protect\citeauthoryear{{Pardo} et~al.,}{{Pardo}
  et~al.}{2016}]{Pardo+2016}
{Pardo} K.,  et~al., 2016, \mn@doi [\apj] {10.3847/0004-637X/831/2/203}, \href
  {http://adsabs.harvard.edu/abs/2016ApJ...831..203P} {831, 203}

\bibitem[\protect\citeauthoryear{{Park}, {Ricotti}, {Natarajan},
  {Bogdanovi{\'c}}  \& {Wise}}{{Park} et~al.}{2016}]{Park+2016}
{Park} K.,  {Ricotti} M.,  {Natarajan} P.,  {Bogdanovi{\'c}} T.,   {Wise}
  J.~H.,  2016, \mn@doi [\apj] {10.3847/0004-637X/818/2/184}, \href
  {http://adsabs.harvard.edu/abs/2016ApJ...818..184P} {818, 184}

\bibitem[\protect\citeauthoryear{{Parkinson}, {Cole}  \& {Helly}}{{Parkinson}
  et~al.}{2008}]{Parkinson+2008}
{Parkinson} H.,  {Cole} S.,   {Helly} J.,  2008, \mn@doi [\mnras]
  {10.1111/j.1365-2966.2007.12517.x}, \href
  {http://adsabs.harvard.edu/abs/2008MNRAS.383..557P} {383, 557}

\bibitem[\protect\citeauthoryear{{Peng}}{{Peng}}{2007}]{Peng2007}
{Peng} C.~Y.,  2007, \mn@doi [\apj] {10.1086/522774}, \href
  {http://adsabs.harvard.edu/abs/2007ApJ...671.1098P} {671, 1098}

\bibitem[\protect\citeauthoryear{{Pezzulli}, {Valiante}  \&
  {Schneider}}{{Pezzulli} et~al.}{2016}]{Pezzulli+2016}
{Pezzulli} E.,  {Valiante} R.,   {Schneider} R.,  2016, \mn@doi [\mnras]
  {10.1093/mnras/stw505}, \href
  {http://adsabs.harvard.edu/abs/2016MNRAS.458.3047P} {458, 3047}

\bibitem[\protect\citeauthoryear{{Planck Collaboration} et~al.,}{{Planck
  Collaboration} et~al.}{2016}]{Planck2016}
{Planck Collaboration} et~al., 2016, \mn@doi [\aap]
  {10.1051/0004-6361/201525830}, \href
  {http://adsabs.harvard.edu/abs/2016A%26A...594A..13P} {594, A13}

\bibitem[\protect\citeauthoryear{{Reines}, {Greene}  \& {Geha}}{{Reines}
  et~al.}{2013}]{Reines+2013}
{Reines} A.~E.,  {Greene} J.~E.,   {Geha} M.,  2013, \mn@doi [\apj]
  {10.1088/0004-637X/775/2/116}, \href
  {http://adsabs.harvard.edu/abs/2013ApJ...775..116R} {775, 116}

\bibitem[\protect\citeauthoryear{{Ricarte} \& {Natarajan}}{{Ricarte} \&
  {Natarajan}}{2018}]{Ricarte&Natarajan2018}
{Ricarte} A.,  {Natarajan} P.,  2018, \mn@doi [\mnras] {10.1093/mnras/stx2851},
  \href {http://adsabs.harvard.edu/abs/2018MNRAS.474.1995R} {474, 1995}

\bibitem[\protect\citeauthoryear{{Rice}, {Lodato}  \& {Armitage}}{{Rice}
  et~al.}{2005}]{Rice+2005}
{Rice} W.~K.~M.,  {Lodato} G.,   {Armitage} P.~J.,  2005, \mn@doi [\mnras]
  {10.1111/j.1745-3933.2005.00105.x}, \href
  {http://adsabs.harvard.edu/abs/2005MNRAS.364L..56R} {364, L56}

\bibitem[\protect\citeauthoryear{{Saglia} et~al.,}{{Saglia}
  et~al.}{2016}]{Saglia+2016}
{Saglia} R.~P.,  et~al., 2016, \mn@doi [\apj] {10.3847/0004-637X/818/1/47},
  \href {http://adsabs.harvard.edu/abs/2016ApJ...818...47S} {818, 47}

\bibitem[\protect\citeauthoryear{{Salvaterra}, {Haardt}, {Volonteri}  \&
  {Moretti}}{{Salvaterra} et~al.}{2012}]{Salvaterra+2012}
{Salvaterra} R.,  {Haardt} F.,  {Volonteri} M.,   {Moretti} A.,  2012, \mn@doi
  [\aap] {10.1051/0004-6361/201219965}, \href
  {http://adsabs.harvard.edu/abs/2012A%26A...545L...6S} {545, L6}

\bibitem[\protect\citeauthoryear{{Sartori}, {Schawinski}, {Treister},
  {Trakhtenbrot}, {Koss}, {Shirazi}  \& {Oh}}{{Sartori}
  et~al.}{2015}]{Sartori+2015}
{Sartori} L.~F.,  {Schawinski} K.,  {Treister} E.,  {Trakhtenbrot} B.,  {Koss}
  M.,  {Shirazi} M.,   {Oh} K.,  2015, \mn@doi [\mnras]
  {10.1093/mnras/stv2238}, \href
  {http://adsabs.harvard.edu/abs/2015MNRAS.454.3722S} {454, 3722}

\bibitem[\protect\citeauthoryear{{Schawinski}, {Koss}, {Berney}  \&
  {Sartori}}{{Schawinski} et~al.}{2015}]{Schawinski+2015}
{Schawinski} K.,  {Koss} M.,  {Berney} S.,   {Sartori} L.~F.,  2015, \mn@doi
  [\mnras] {10.1093/mnras/stv1136}, \href
  {http://adsabs.harvard.edu/abs/2015MNRAS.451.2517S} {451, 2517}

\bibitem[\protect\citeauthoryear{{Sesana}, {Haardt}, {Madau}  \&
  {Volonteri}}{{Sesana} et~al.}{2005}]{Sesana+2005}
{Sesana} A.,  {Haardt} F.,  {Madau} P.,   {Volonteri} M.,  2005, \mn@doi [\apj]
  {10.1086/428492}, \href {http://adsabs.harvard.edu/abs/2005ApJ...623...23S}
  {623, 23}

\bibitem[\protect\citeauthoryear{{Sesana}, {Volonteri}  \& {Haardt}}{{Sesana}
  et~al.}{2007}]{Sesana+2007}
{Sesana} A.,  {Volonteri} M.,   {Haardt} F.,  2007, \mn@doi [\mnras]
  {10.1111/j.1365-2966.2007.11734.x}, \href
  {http://adsabs.harvard.edu/abs/2007MNRAS.377.1711S} {377, 1711}

\bibitem[\protect\citeauthoryear{{Shankar}, {Weinberg}  \&
  {Miralda-Escud{\'e}}}{{Shankar} et~al.}{2009}]{Shankar+2009}
{Shankar} F.,  {Weinberg} D.~H.,   {Miralda-Escud{\'e}} J.,  2009, \mn@doi
  [\apj] {10.1088/0004-637X/690/1/20}, \href
  {http://adsabs.harvard.edu/abs/2009ApJ...690...20S} {690, 20}

\bibitem[\protect\citeauthoryear{{Somerville}}{{Somerville}}{2009}]{Somerville2009}
{Somerville} R.~S.,  2009, \mn@doi [\mnras] {10.1111/j.1365-2966.2009.15325.x},
  \href {https://ui.adsabs.harvard.edu/#abs/2009MNRAS.399.1988S} {399, 1988}

\bibitem[\protect\citeauthoryear{{Stacy}, {Bromm}  \& {Lee}}{{Stacy}
  et~al.}{2016}]{Stacy+2016}
{Stacy} A.,  {Bromm} V.,   {Lee} A.~T.,  2016, \mn@doi [\mnras]
  {10.1093/mnras/stw1728}, \href
  {http://adsabs.harvard.edu/abs/2016MNRAS.462.1307S} {462, 1307}

\bibitem[\protect\citeauthoryear{{Stone} \& {Metzger}}{{Stone} \&
  {Metzger}}{2016}]{Stone&Metzger2016}
{Stone} N.~C.,  {Metzger} B.~D.,  2016, \mn@doi [\mnras]
  {10.1093/mnras/stv2281}, \href
  {http://adsabs.harvard.edu/abs/2016MNRAS.455..859S} {455, 859}

\bibitem[\protect\citeauthoryear{{Stone}, {K{\"u}pper}  \& {Ostriker}}{{Stone}
  et~al.}{2017}]{Stone+2017}
{Stone} N.~C.,  {K{\"u}pper} A.~H.~W.,   {Ostriker} J.~P.,  2017, \mn@doi
  [\mnras] {10.1093/mnras/stx097}, \href
  {http://adsabs.harvard.edu/abs/2017MNRAS.467.4180S} {467, 4180}

\bibitem[\protect\citeauthoryear{{Treister} et~al.,}{{Treister}
  et~al.}{2009}]{Treister+2009}
{Treister} E.,  et~al., 2009, \mn@doi [\apj] {10.1088/0004-637X/706/1/535},
  \href {http://adsabs.harvard.edu/abs/2009ApJ...706..535T} {706, 535}

\bibitem[\protect\citeauthoryear{{Treister}, {Schawinski}, {Urry}  \&
  {Simmons}}{{Treister} et~al.}{2012}]{Treister+2012}
{Treister} E.,  {Schawinski} K.,  {Urry} C.~M.,   {Simmons} B.~D.,  2012,
  \mn@doi [\apjl] {10.1088/2041-8205/758/2/L39}, \href
  {http://adsabs.harvard.edu/abs/2012ApJ...758L..39T} {758, L39}

\bibitem[\protect\citeauthoryear{{Treister}, {Schawinski}, {Volonteri}  \&
  {Natarajan}}{{Treister} et~al.}{2013}]{Treister+2013}
{Treister} E.,  {Schawinski} K.,  {Volonteri} M.,   {Natarajan} P.,  2013,
  \mn@doi [\apj] {10.1088/0004-637X/778/2/130}, \href
  {http://adsabs.harvard.edu/abs/2013ApJ...778..130T} {778, 130}

\bibitem[\protect\citeauthoryear{{Tremaine} et~al.,}{{Tremaine}
  et~al.}{2002}]{Tremaine+2002}
{Tremaine} S.,  et~al., 2002, \mn@doi [\apj] {10.1086/341002}, \href
  {http://adsabs.harvard.edu/abs/2002ApJ...574..740T} {574, 740}

\bibitem[\protect\citeauthoryear{{Tremmel}, {Governato}, {Volonteri}, {Quinn}
  \& {Pontzen}}{{Tremmel} et~al.}{2017}]{Tremmel+2017}
{Tremmel} M.,  {Governato} F.,  {Volonteri} M.,  {Quinn} T.~R.,   {Pontzen} A.,
   2017, preprint, \href {http://adsabs.harvard.edu/abs/2017arXiv170807126T} {}
  (\mn@eprint {arXiv} {1708.07126})

\bibitem[\protect\citeauthoryear{{Ueda}, {Akiyama}, {Hasinger}, {Miyaji}  \&
  {Watson}}{{Ueda} et~al.}{2014}]{Ueda+2014}
{Ueda} Y.,  {Akiyama} M.,  {Hasinger} G.,  {Miyaji} T.,   {Watson} M.~G.,
  2014, \mn@doi [\apj] {10.1088/0004-637X/786/2/104}, \href
  {http://adsabs.harvard.edu/abs/2014ApJ...786..104U} {786, 104}

\bibitem[\protect\citeauthoryear{{Villforth} et~al.,}{{Villforth}
  et~al.}{2017}]{Villforth+2017}
{Villforth} C.,  et~al., 2017, \mn@doi [\mnras] {10.1093/mnras/stw3037}, \href
  {http://adsabs.harvard.edu/abs/2017MNRAS.466..812V} {466, 812}

\bibitem[\protect\citeauthoryear{{Volonteri} \& {Bellovary}}{{Volonteri} \&
  {Bellovary}}{2012}]{Volonteri&Bellovary2012}
{Volonteri} M.,  {Bellovary} J.,  2012, \mn@doi [Reports on Progress in
  Physics] {10.1088/0034-4885/75/12/124901}, \href
  {http://adsabs.harvard.edu/abs/2012RPPh...75l4901V} {75, 124901}

\bibitem[\protect\citeauthoryear{{Volonteri} \& {Natarajan}}{{Volonteri} \&
  {Natarajan}}{2009}]{Volonteri&Natarajan2009}
{Volonteri} M.,  {Natarajan} P.,  2009, \mn@doi [\mnras]
  {10.1111/j.1365-2966.2009.15577.x}, \href
  {http://adsabs.harvard.edu/abs/2009MNRAS.400.1911V} {400, 1911}

\bibitem[\protect\citeauthoryear{{Volonteri} \& {Rees}}{{Volonteri} \&
  {Rees}}{2005}]{Volonteri&Rees2005}
{Volonteri} M.,  {Rees} M.~J.,  2005, \mn@doi [\apj] {10.1086/466521}, \href
  {http://adsabs.harvard.edu/abs/2005ApJ...633..624V} {633, 624}

\bibitem[\protect\citeauthoryear{{Warren}, {Quinn}, {Salmon}  \&
  {Zurek}}{{Warren} et~al.}{1992}]{Warren+1992}
{Warren} M.~S.,  {Quinn} P.~J.,  {Salmon} J.~K.,   {Zurek} W.~H.,  1992,
  \mn@doi [\apj] {10.1086/171937}, \href
  {http://adsabs.harvard.edu/abs/1992ApJ...399..405W} {399, 405}

\bibitem[\protect\citeauthoryear{{Weigel}, {Schawinski}, {Treister},
  {Trakhtenbrot}  \& {Sanders}}{{Weigel} et~al.}{2018}]{Weigel+2018}
{Weigel} A.~K.,  {Schawinski} K.,  {Treister} E.,  {Trakhtenbrot} B.,
  {Sanders} D.~B.,  2018, \mn@doi [\mnras] {10.1093/mnras/sty383}, \href
  {http://adsabs.harvard.edu/abs/2018MNRAS.tmp..373W} {}

\bibitem[\protect\citeauthoryear{{Wu} et~al.,}{{Wu} et~al.}{2015}]{Wu+2015}
{Wu} X.-B.,  et~al., 2015, \mn@doi [\nat] {10.1038/nature14241}, \href
  {http://adsabs.harvard.edu/abs/2015Natur.518..512W} {518, 512}

\bibitem[\protect\citeauthoryear{{Yue}, {Ferrara}, {Salvaterra}, {Xu}  \&
  {Chen}}{{Yue} et~al.}{2013}]{Yue+2013}
{Yue} B.,  {Ferrara} A.,  {Salvaterra} R.,  {Xu} Y.,   {Chen} X.,  2013,
  \mn@doi [\mnras] {10.1093/mnras/stt826}, \href
  {http://adsabs.harvard.edu/abs/2013MNRAS.433.1556Y} {433, 1556}

\bibitem[\protect\citeauthoryear{{den Brok} et~al.,}{{den Brok}
  et~al.}{2015}]{DenBrok+2015}
{den Brok} M.,  et~al., 2015, \mn@doi [\apj] {10.1088/0004-637X/809/1/101},
  \href {http://adsabs.harvard.edu/abs/2015ApJ...809..101D} {809, 101}

\bibitem[\protect\citeauthoryear{{van den Bosch}}{{van den
  Bosch}}{2016}]{vandenBosch2016}
{van den Bosch} R.~C.~E.,  2016, \mn@doi [\apj] {10.3847/0004-637X/831/2/134},
  \href {http://adsabs.harvard.edu/abs/2016ApJ...831..134V} {831, 134}

\makeatother
\end{thebibliography}

\end{document}